\documentclass[12pt,journal,onecolumn]{IEEEtran}
\usepackage{amsmath, epsfig, cite}
\usepackage{amssymb}
\usepackage{amsfonts}
\usepackage{latexsym}
\usepackage{longtable}
\usepackage{tabularx}
\usepackage{color}
\usepackage{multirow}
\usepackage{pgf}
\usepackage{tikz}
\usetikzlibrary{arrows, automata, positioning, calc}
\usepackage{array}
\usepackage{colortbl}
\usepackage{comment}

\usepackage{graphicx}
\usepackage{makecell}
\usepackage{diagbox}

\newcommand{\field}[1]{\mathbb{#1}}
\newcommand{\F}{\field{F}}

\newcommand{\cC}{{\cal C}}

\newcommand{\cS}{{\cal S}}
\newcommand{\cT}{{\cal T}}

\newcommand{\cR}{{\cal R}}

\newcommand{\cH}{{\cal H}}

\newcommand{\bG}{\mathbf{G}}
\newcommand{\bv}{\mathbf{v}}
\newcommand{\bu}{\mathbf{u}}
\newcommand{\bw}{\mathbf{w}}
\newcommand{\bz}{\mathbf{z}}
\newcommand{\by}{\mathbf{y}}
\newcommand{\bx}{\mathbf{x}}
\newcommand{\be}{\mathbf{e}}
\newcommand{\bL}{\mathbf{L}}
\newcommand{\bA}{\mathbf{A}}
\newcommand{\bB}{\mathbf{B}}

\newcommand{\bI}{\mathbf{I}}
\newcommand{\bU}{\mathbf{U}}
\newcommand{\bH}{\mathbf{H}}
\newcommand{\bzero}{\mathbf{0}}

\newcommand{\supp}{\text{supp}}

\newtheorem{theorem}{Theorem}
\newtheorem{proposition}[theorem]{Proposition}

\newtheorem{corollary}[theorem]{Corollary}
\newtheorem{definition}[theorem]{Definition}
\newtheorem{lemma}[theorem]{Lemma}
\newtheorem{construction}{Construction}
\newtheorem{conjecture}[theorem]{Conjecture}
\newtheorem{example}{Example}
\newtheorem{remark}{Remark}
\newcommand{\qed}{{\hfill\rule{4pt}{7pt}}}

\newcommand{\tabincell}[2]{\begin{tabular}{@{}#1@{}}#2\end{tabular}}

\begin{document}

\title{\textbf{Bounds on the Length\\
of Functional PIR and Batch codes}}

%\author{\IEEEauthorblockN{Yiwei Zhang, Eitan Yaakobi, and Tuvi Etzion}\\
%\IEEEauthorblockA{Department of Computer Science\\
%Technion---Israel Institute of Technology, Haifa, Israel\\
%\emph{\{ywzhang, yaakobi, etzion\}@cs.technion.ac.il}
%}}

\author{
{\small Yiwei Zhang, \hspace{0.1cm}
Tuvi Etzion,~\IEEEmembership{\small Fellow,~IEEE}, \hspace{0.1cm}
Eitan Yaakobi,~\IEEEmembership{\small Senior Member,~IEEE}}
\thanks{The authors are with the Department of Computer Science, Technion -- Israel Institute of Technology, Haifa 3200003, Israel, (e-mail: \{ywzhang,etzion,yaakobi\}@cs.technion.ac.il).}
\thanks{The material in this paper will be presented in part at the
IEEE International Symposium on Information
Theory (ISIT 2019), Paris, France, July 2019.}%
\thanks{Eitan Yaakobi and Yiwei Zhang were supported in part by the ISF grant 1817/18;
Tuvi Etzion and Yiwei Zhang were supported in part by the BSF-NSF grant 2016692;
Yiwei Zhang was also supported in part by a Technion Fellowship.}
}

\maketitle

\begin{abstract}
A \emph{functional $k$-PIR code} of dimension $s$ consists of $n$ servers storing linear combinations
of $s$ linearly independent information symbols. Any linear combination of the $s$ information
symbols can be recovered by $k$ disjoint subsets of servers.
The goal is to find
the smallest number of servers for given $k$ and $s$. We provide lower bounds on the number
of servers and constructions which yield upper bounds on this number. For $k \leq 4$, exact bounds
on the number of servers are proved. Furthermore, we provide some asymptotic bounds. The problem coincides with the well known
private information retrieval problem based on a coded database to reduce the storage overhead,
when each linear combination contains exactly one information symbol.

If any multiset of size $k$ of linear combinations from the
linearly independent information symbols can be recovered
by $k$ disjoint subset of servers, then the servers form a \emph{functional $k$-batch code}.
A~functional $k$-batch code is a functional $k$-PIR code, where all
the $k$ linear combinations in the multiset are equal.
We provide some bounds on the number of servers for functional $k$-batch codes.
In particular we present a random construction and a construction based on
simplex codes, WOM codes, and RIO codes.
\end{abstract}

\section{Introduction}
\label{sec:intro}

\subsection{General Background}

A Private Information Retrieval (PIR) protocol allows a user to retrieve
a data item from a database, in such a way that the servers storing the data
will get no information about which data item was retrieved. The problem was introduced in~\cite{CGKS98}.
The protocol to achieve this goal assumes that the servers are curious but honest,
so they don't collude. It is also assumed that the database is error-free
and is synchronized all the time. For a set of $k$ servers, the goal is to design an
efficient $k$-server PIR protocol, where efficiency is measured by the total
number of bits transmitted by all parties involved. This model is called
\emph{information-theoretic} PIR; there is also \emph{computational}
PIR, in which the privacy is defined in terms of the inability of a server to compute
which item was retrieved in a reasonable time~\cite{KuOs97}. We continue to consider
only the information-theoretic PIR.

The classic model of PIR assumes that each server stores
a copy of an $s$-bit database, so the \emph{storage overhead}, namely the ratio between the total
number of bits stored by all servers and the size of the
database, is~$k$. However, recent work combines PIR protocols
with techniques from distributed storage (where each server stores only a coded fraction of the database)
to reduce the storage overhead. This approach was first considered in~\cite{SRR14},
and several papers have developed this direction
further, e.g.~\cite{ALS14,CHY14,CHY15}.
Our discussion on PIR will follow the breakthrough approach presented
in~\cite{FVY15,FVY15a}, which shows that $n$ servers
(for some $n>k$) may emulate a $k$-server PIR protocol with storage overhead significantly lower than $k$.
The scheme used for this purpose is called a \emph{$k$-PIR} and will be discussed in the next paragraph.

The $s$-bit database $\cS$ is considered as the information bits of a linear code of length $n$
and dimension $s$. This code has an $s \times n$ generator matrix $\bG$.
The linear combinations related to the codeword $\cS \bG$ are stored in the $n$ servers.
In other words, the $i$-th server stores the linear combination generated when the $s$-bit
information word is multiplied by the $i$-th column of $\bG$. The generator matrix $\bG$ represents
a $k$-PIR scheme if there are $k$ pairwise disjoint subsets of $[n] \triangleq \{1,2,\ldots,n\}$, $R_1,R_2,\ldots,R_k$,
such that the sum of the columns of $\bG$ related to each such subset is the data item
(out of the $s$ data items) which the user wants to retrieve.
Using these $k$ subsets any known $k$-PIR protocol can be emulated with the
given $n$ servers. The advantage of this scheme is a smaller amount
of storage used for a $k$-PIR protocol. The goal in the design of such a PIR scheme is to find the smallest $n$,
given $s$ and~$k$. This problem was considered in several papers, e.g.~\cite{AsYa18,FVY15,FVY15a,LiRo17,RaVa16,VRV17,Wooters16}.

In all the PIR protocols known in the literature, the user wants to retrieve one out of the
$s$ information bits of the database. As will be described in the sequel, PIR codes and their
generalizations are similar to other concepts in coding theory. For example, there is also the similar requirement for codes
with availability~\cite{RPDV16}, which are important in applications of distributed storage codes.
In some of the related applications, it is quite natural that it will be required to
retrieve a linear combination of the $s$ bits of information symbols. Such a scheme will be called
a \emph{k-functional PIR code} (this is some abuse of definition since for the private information retrieval
application such a retrieval of linear combinations is
not required). Given $s$ and $k$ we would like to find the smallest $n$ for which
a functional $k$-PIR exists. This is one of the two targets of the current paper.

The definition of a $k$-PIR code appears to be a special case of a $k$-batch code. The concept of a batch scheme
was first proposed by Ishai et al.~\cite{IKOS04}, which was motivated by different applications for load-balancing
in storage and cryptographic protocols. Originally, batch codes were defined in a very general form, i.e., $s$ information symbols
are encoded into $n$-tuples of strings where each string is called a bucket.
Each bucket contains a few linear combinations of the information symbols.
A single user wants to retrieve a batch of $k$ distinct
data items (out of the $s$ data items) by reading at most $t$ symbols from each bucket. The goal in the design of a batch scheme is to
find the smallest total length of all the buckets, given $s$, $k$, $t$ and $n$.

A stronger variant of batch codes~\cite{IKOS04} is intended for a multi-user application instead of
a single-user setting, known as the \emph{multiset batch codes}. In this variant we have $k$ different users each requesting a data item,
where some of the requests are allowed to be the same. Therefore all the $k$ requests constitute a multiset of data items
(each being one out of the $s$ data items, replications allowed). Moreover, each bucket is allowed to be accessed by at most one user.
A special case of a multiset batch code is when each bucket contains only one symbol.
This model is called a \emph{primitive multiset batch code} \cite{IKOS04} (or a \emph{$k$-batch code} in short)
and it is a family of batch codes that was most studied in the literature.
In the rest of this paper, we restrict our definition of batch codes only to primitive multiset batch codes.
Similarly as for a PIR code, a batch code is represented by an $s \times n$ generator matrix $\bG$. It is a $k$-batch scheme if there are $k$
pairwise disjoint subsets of $[n]$, $R_1,R_2,\ldots,R_k$, such that the $k$ sums from
each subset of the columns in $\bG$ constitute a multiset of data items which some $k$ users want to retrieve.
%these $k$ sums of the columns of $\bG$ related to each such subset constitute a multiset of data items which some $k$ users want to retrieve.
Hence, the requests in a $k$-PIR are a special case of the requests in a $k$-batch
when the multiset contains only one specific item $k$ times. Therefore a $k$-batch code can always work as a $k$-PIR code but not vice versa.
The goal in the design of a batch scheme is to find the smallest $n$, given $s$ and~$k$.
This problem was considered in several papers, e.g.~\cite{AsYa18,BCSY18,IKOS04,RSGA16,VaYa16}.

Similarly as our generalization of PIR into functional PIR, by setting the requests to be a multiset of linear
combinations of the $s$ bits of information symbols, a batch code is generalized into a \emph{functional batch code}.
Given $s$ and $k$ we would like to find the smallest $n$ for which a functional $k$-batch code exists.
This is the second target of the current paper.

A special case of batch codes, called \emph{switch codes}, were recently studied for
network applications~\cite{BCSY18,CGTZ15,WKC15,WKCB17,WSCB13}. This family of codes was
first proposed by Wang et al.~\cite{WSCB13} and these codes were designed to increase the parallelism
of data writing and reading processes in network switches. A network switch is required
to write $n$ incoming packets and read $k$ outgoing packets while using $m$ memory banks,
each able to write and read one packet per time unit. Each set of $n$ packets written to
the switch simultaneously is called a generation. The objective is to store the packets
in the banks such that every request of $k$ packets, which can be from previous generations,
can be handled by reading at most one packet from each bank. Even though batch codes
and switch codes were proved to be equivalent~\cite{BCSY18}, switch codes are commonly
designed for the special case of $k=n$, which balances the output and input switching rates.

A related family of codes to functional batch codes is the
family of \emph{random I/O (RIO) codes}. This family of codes was
recently introduced by Sharon and Alrod~\cite{SA13} and provides a coding scheme to improve
the random input/output performance of flash memories. An $(n,M,t)$ RIO code stores $t$ pages
in $n$ cells with $t+1$ levels such that it is enough to sense a single read threshold in
order to read any of the $t$ pages. Sharon and Alrod showed in~\cite{SA13} that the design of RIO codes is equivalent to the design of \emph{write-once memory} (\emph{WOM}) \emph{codes}~\cite{CGM86,G87,RS82,YKSVW12}. The latter family of codes attracted substantial attention in recent years in order to improve the lifetime of flash memories by allowing writing multiple messages to the memory without the need for an erase operation. However, while in WOM codes, the messages are received one after the other and thus are not known in advance, in RIO codes the information of all logical pages can be known in advance when programming the cells. This variant of RIO codes, called \emph{parallel RIO codes}, was introduced in~\cite{YM16}. A recent construction of parallel RIO codes~\cite{YKL17} used the coset coding scheme~\cite{CGM86} with Hamming codes in order to construct parallel RIO codes. In fact, this construction is equivalent to the requirements of functional batch codes, and thus every functional batch code can be used as a parallel RIO code as well. The other direction does not necessarily hold since parallel RIO codes do not have to be linear, as opposed to functional batch codes. The codes from~\cite{YKL17} gave two constructions of functional batch codes (which are parallel RIO codes) with the following parameters: $(s=3,k=4,n=7)$ and $(s=4,k=8,n=15)$.

\subsection{General Description of the Problem}
\label{sec:pre}

Assume there are $n$ servers, each storing a linear combination of $s$ linearly independent items.
Each of these $s$ items will be called an \emph{information symbol}. Each linear combination which consists of at least
one of these information symbols will be called a \emph{coded symbol}.
There are $k$ users who want to retrieve $k$ linear combinations of items from these servers.
Each such linear combination which a user wants to retrieve will be called a \emph{request}.
Each user has exactly one such request and he should approach
a set of servers to obtain his request. The set of servers which are approached by two different users
must be disjoint. We would like to know the smallest number of servers which is required
to satisfy any $k$ requests of the $k$~users. This scheme
will be called a \emph{functional $k$-batch code} (functional $k$-batch for short, and similarly done
for the related concepts). If each request contains
exactly one information symbol, then the scheme will be called a \emph{$k$-batch code}.

If the $k$ requests are the same (linear combination) then the scheme will be called a \emph{functional $k$-PIR code}
and furthermore if these $k$ requests contain the same information symbol, then the scheme will be called a \emph{$k$-PIR code}.
This definition for $k$-PIR coincides with the definition for $k\text{-PIR}$ given in~\cite{FVY15,FVY15a} for a single user.
Let $FB(s,k)$ ($B(s,k)$, $FP(s,k)$, $P(s,k)$, respectively) be the minimum number of servers required for $s$ items
and $k$ requests for functional $k$-batch ($k$-batch, functional $k$-PIR, $k$-PIR, respectively). Next, we present the formal definition for
functional $k$-batch code ($k$-batch code, functional $k$-PIR code, $k$-PIR code, respectively).

%A \emph{functional $k$-PIR code} of length $n$ and dimension $s$ consists of $n$ servers and $s$ information symbols $\{x_1,x_2,\ldots,x_s\}$.
%Each server store a nontrivial linear combination of the information symbols, i.e. the $j$-th server stores
%a linear combination $Y_j$. For any request of a linear combination of the information symbols $Z$, there are
%$k$ pairwise disjoint subsets $R_1,R_2,\ldots,R_k$ of $[n] \triangleq \{ 1,2,\ldots,n\}$ such that the sum of the linear combinations
%in the related servers of $R_j$, $1 \leq j \leq k$, is $Z$, i.e. $\sum_{\ell \in R_j} Y_\ell = Z$. The functional $k$-PIR code
%can be also represented by an $s \times n$ matrix $\bG$ in which the $j$-th column has \emph{ones} in positions $i_1, i_2,\ldots,i_\ell$
%if and only if the $j$-server has the linear combination $x_{i_1} + x_{i_2} + \cdots + x_{i_\ell}$.

A \emph{functional $k$-batch code} of length $n$ and dimension $s$ consists
of $n$ servers and $s$ information symbols $\{x_1,x_2,\ldots,x_s\}$.
Each server stores a nontrivial linear combination of the information symbols
(which are the coded symbols), i.e. the $j$-th server stores
a linear combination $Y_j$, $1\le j \le n$. For any request of $k$ linear combinations $\bv_1,\bv_2,\ldots,\bv_k$
(not necessarily distinct) of the information symbols, there are
$k$ pairwise disjoint subsets $R_1,R_2,\ldots,R_k$ of $[n]$ such that the sum of the linear combinations
in the related servers of $R_j$, $1 \leq j \leq k$, is $\bv_j$, i.e. $\sum_{\ell \in R_j} Y_\ell = \bv_j$.
Each such $\bv_i$ will be called a \emph{requested symbol} and
each such subset $R_j$ will be called a \emph{recovery set}. A functional $k$-batch code
can be also represented by an $s \times n$ matrix $\bG$ in which the $j$-th column has \emph{ones} in positions $i_1, i_2,\ldots,i_\ell$
if and only if the $j$-th server stores the linear combination $x_{i_1} + x_{i_2} + \cdots + x_{i_\ell}$.

To summarize, a \emph{$k$-batch code} is defined similarly to a functional $k$-batch code,
where each one of the requests $\bv_1,\bv_2,\ldots,\bv_k$ contains
exactly one information symbol.
A \emph{functional $k$-PIR code} is defined similarly to a functional $k$-batch code,
where all the $\bv_i$'s equal to one linear combination~$\bv$.
A \emph{$k$-PIR code} is defined similarly to a functional $k$-PIR code, where the linear combination $\bv$ contains
exactly one information symbol.

By these definitions, a (functional) batch code is also a (functional) PIR code (where all the requests are equal)
and a functional batch (PIR, respectively) code is also a batch (PIR, respectively) code, but not vice versa.
Thus, we have the following relationships among these four families of codes.
\smallskip

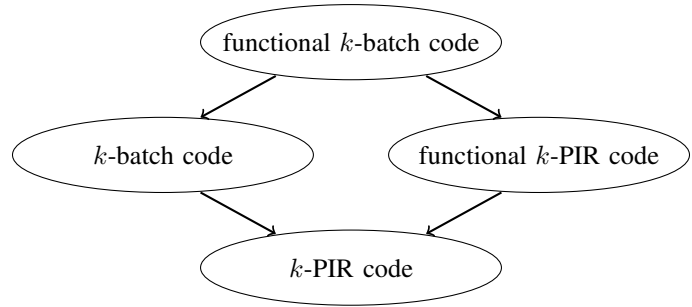
\begin{figure*}[!h]
\centering
\begin{tikzpicture}[scale=0.5]
     \tikzstyle{edge} = [->,draw,thick,black]
     \draw (0,3) ellipse [x radius=4, y radius=1];
     \draw (0,-3) ellipse [x radius=4, y radius=1];
     \draw (-5,0) ellipse [x radius=4, y radius=1];
     \draw (5,0) ellipse [x radius=4, y radius=1];

     \begin{small}\node at (0,3)  {functional $k$-batch code};\end{small}
     \begin{small}\node at (-5,0)  {$k$-batch code};\end{small}
     \begin{small}\node at (5,0)  {functional $k$-PIR code};\end{small}
     \begin{small}\node at (0,-3)  {$k$-PIR code};\end{small}

     \draw[edge] (-2,2.1)-- (-4,1);
     \draw[edge] (2,2.1)-- (4,1);
     \draw[edge] (-4,-1)-- (-2,-2.1);
     \draw[edge] (4,-1)-- (2,-2.1);

\end{tikzpicture}
\end{figure*}

\vspace{-0.7cm}

\subsection{Basic Results}
\label{sec:basic}

Our goal in this paper is to obtain lower and upper bounds on
$FB(s,k)$ and $FP(s,k)$, since relatively good bounds on $B(s,k)$ and $P(s,k)$
are known from the literature.
Some of these bounds on $B(s,k)$ and $P(s,k)$ were derived in~\cite{AsYa18,BCSY18,FVY15,LiCo04,RSGA16,RaVa16,VaYa16,Wooters16} and are
summarized as follows.
\begin{lemma}\label{lem:known}
$~$
\begin{enumerate}
\item For each $s \geq 1$, $P(s,2^{s-1})=2^s -1$~\cite{FVY15a}.

\item For each $s \geq 1$, $B(s,2^{s-1})=2^s -1$~\cite{WSCB13}.

\item When $k$ is a fixed integer, $P(s,k)=s+\Theta(\sqrt{s})$~\cite{FVY15,RaVa16,Wooters16}.

\item $B(s,k)=s+\Theta(\sqrt{s})$ for $k=3,4,5$~\cite{VaYa16,AsYa18}.

\item $B(s,k)=s+O(\sqrt{s}\log{s})$ for $k\ge6$~\cite{VaYa16}.

\item $B(s,s^{1/3})\le 2s$ \cite{RSGA16}.

\item $B(s,s^{\varepsilon})\le s+s^{7/8}$ for $7/32\le \varepsilon \le1/4$ \cite{RSGA16}.

\item $B(s,s^{\varepsilon})\le s+s^{4\varepsilon}$ for $1/5< \varepsilon \le 7/32$ \cite{RSGA16}.

\item $B(s,s)\le 2s^{1.5}$ \cite{BCSY18}.

\item $P(s,\sqrt{s})=s+O(s^{(\log3/2)})$ \cite{LiCo04}.

\item $P(s,s^{\varepsilon})=s+O(s^{0.5+\varepsilon})$, $0 < \epsilon < 1/2$~\cite{LiCo04}.

\item $B(s,k=\Theta(s^{\varepsilon}))=s+o(s)$, $0 < \epsilon < 1$~\cite{AsYa18}.

\item $P(s,k=\Theta(s^{\varepsilon}))=s+o(s)$, $0 < \epsilon < 1$~\cite{AsYa18}.

\item $B(s,k=o(n^{1/3}/ \log n))=s+O(k^{3/2}\sqrt{n}\log n)$~\cite{PoVo19}.

\item For $k<\frac{1}{\ell^2}n^{1/(2\ell+1)}$, $\ell$ is a positive integer,
$B(s,k)=s+O(kn^{\frac{\ell+1}{2\ell+1}})$~\cite{PoVo19}.
\end{enumerate}
\end{lemma}

For a binary vector $\bv$, let $\supp(\bv)$ denote the support of $\bv$, i.e., the
set of nonzero entries of~$\bv$.
Some simple bounds on $FB(s,k)$ and on $FP(s,k)$ are derived in the following theorem.

\begin{theorem}
\label{thm:trivial}
If $s$ and $k$ are positive integers, then
\begin{enumerate}
\item For $k>1$, $FB(s,k) > FB(s,k-1)$.

\item For $k>1$, $FP(s,k) > FP(s,k-1)$.

\item For $s \geq 1$, $FP(s,1)=FB(s,1)=s$.

\item For $s \geq 1$, $FP(s,2)=s+1$.

\item For $s \geq 1$ and $k \geq 1$, $FP(s,2k)=FP(s,2k-1)+1$.

\item For $s \geq 1$ and $k \geq 1$, $FP(s,k)\le FB(s,k)$.
\end{enumerate}
\end{theorem}

\begin{IEEEproof}
\begin{enumerate}
\item If any server is removed from a $k$-batch code then the remaining servers
form a $(k-1)$-batch code and hence $FB(s,k) > FB(s,k-1)$ for $k > 1$.

\item If any server is removed from a $k$-PIR code then the remaining servers
form a $(k-1)$-PIR code and hence $FP(s,k) > FP(s,k-1)$ for $k>1$.

\item If $FP(s,1)<s$ or $FB(s,1)<s$ then the rank of the information stored by the symbols
is less than $s$ and hence there is a linear combination not in their spanned information
that cannot be recovered, a contradiction. Hence, $FP(s,1)\geq s$ and $FB(s,1) \geq s$.
An 1-PIR code (1-batch code) of length $s$ is constructed by storing
the information symbol~$x_j$, $1\le j \le s$, in the $j$-th server. Therefore,
$FP(s,1)\leq s$ and $FB(s,1) \leq s$ and the claim follows.

\item Since $FP(s,k) > FP(s,k-1)$ for $k>1$, it follows that $FP(s,2) \geq s+1$.
Consider the code of length $s+1$, where the $j$-th server stores the information symbol $x_j$, $1\le j \le s$ and
the $(s+1)$-th server stores a parity symbol $\sigma=\sum x_j$. For any requested symbol $\mathbf{v}$,
let $\supp(\bv)$ be its support set. The requested symbol $\mathbf{v}$ can be recovered
from the servers indexed by its support set and simultaneously by the
remaining servers, since the sum of the symbols from all servers is zero, i.e., $\mathbf{v}=\sum_{j\in \supp(\mathbf{v})} x_j=\sum_{j\notin \supp(\mathbf{v})} x_j + \sigma$.

\item From the previous parts of the theorem we have
$FP(s,2k-1)\le FP(s,2k)-1$. On the other hand, suppose we have a functional $(2k-1)$-PIR code
with $FP(s,2k-1)$ servers. Add a server storing a global parity symbol, i.e., the sum of the symbols
in the other servers.
Any requested symbol can be recovered $(2k-1)$ times in the same way as in
the functional $(2k-1)$-PIR code. It can be recovered one additional time
by using all the remaining servers, since the global parity implies that the sum of the symbols
from all servers is zero. This implies that $FP(s,2k) \leq FP(s,2k-1)+1$ and thus, $FP(s,2k)=FP(s,2k-1)+1$.

\item Follows from the observation that a functional $k$-batch code can serve as a functional $k$-PIR code.
\end{enumerate}
\end{IEEEproof}

Another basic result concerning PIR and batch codes with $s$ information symbols and $n$ servers
is related to their presentation via a binary $s \times n$ matrix $\bG$ whose columns represent the information
in the servers. In other words, the entries on the $i$-th column
of $\bG$ have \emph{ones} which relate to the information
symbols used in the coded symbol stored in the $i$-th server. A code in which each information symbol is stored
in a server will be called \emph{systematic}. An intriguing question is whether for
all PIR codes and/or batch codes there are related systematic codes
with the same parameters? We conjecture that this is indeed the case, but there is
no proof for this property for $k$-PIR and $k$-batch and it is left as an open problem. We can solve this question
in the case of functional PIR and functional batch.

\begin{lemma} \label{lem:systematic}
If there exists a functional $k$-PIR (batch) code $\cC$ of length $n$ and dimension $s$,
then there exists a systematic functional $k$-PIR (batch) code of length $n$ and dimension $s$.
\end{lemma}
\begin{IEEEproof}
Assume first that $\cC$ is a functional $k$-PIR code that is represented by
an $s \times n$ matrix $\bG$. If $rank(\bG)<s$, then there exists a nonzero
vector $\bv$ not in the column space of $\bG$ which cannot be recovered, a contradiction.
Therefore, $rank(\bG)=s$. Assume w.l.o.g. that $\bG=[\bA ~\bB ]$, where $\bA$ is an $s \times s$ matrix,
$\bB$ is an $s \times (n-s)$ matrix, and $rank(\bA)=s$, i.e., $A$ is an invertible matrix. We claim
that $\bG'=[\bA^{-1}\bA~~\bA^{-1}\bB]$ is also a matrix representing a functional $k$-PIR code $\cC'$.
For each request~$\bv$ (for the code $\cC'$), consider how $\bA\bv$ is recovered $k$ times using $\cC$. For any set of columns in $\bG$ summing
up to~$\bA \bv$, we use the columns in $\bG'$ with the same indices. These columns sum to $\bA^{-1}\bA \bv=\bv$.
Therefore, a systematic functional $k$-PIR code of length $n$ and dimension $s$ is obtained.

A similar proof works if $\cC$ is a functional $k$-batch code.
\end{IEEEproof}

Some more simple bounds on $FP(s,k)$ are given in the following theorem.
\begin{theorem}
\label{thm:simple}
If $s,t,s_1,s_2,k_1,k_2$ are positive integers, then
\begin{enumerate}
\item[(1)] $FP(s,2^{s-1}) = 2^s-1$.

\item[(2)] $FP(s,k_1+k_2) \leq FP(s,k_1)+FP(s,k_2)$.

\item[(3)] $FP(s_1+s_2,k) \leq FP(s_1,k)+FP(s_2,k)$.

\item[(4)] $FP(rt,2^r) \leq 2t(2^r-1)$.
\end{enumerate}
\end{theorem}
\begin{IEEEproof}
\begin{enumerate}
\item[(1)] By Lemma~\ref{lem:known}(1), we have that $FP(s,2^{s-1}) \geq P(s,2^{s-1}) = 2^s-1$, so we only
need to show that $FP(s,2^{s-1}) \leq 2^s-1$. Indeed, a functional $2^{s-1}$-PIR code is obtained from
an $s \times (2^s-1)$ matrix whose columns are all the columns of length $s$. Each request $\bv$ can
be recovered $2^{s-1}$ times, by $2^{s-1}-1$ pairs $(\bu,\bu+\bv)$ and by $\bv$ itself.

\item[(2)] This result follows immediately by concatenating the matrices which represent the
functional $k_1$-PIR code and the functional $k_2$-PIR code with $s$ information symbols.

\item[(3)] Assume $A$ and $B$ are the matrices which represent the functional $k$-PIR codes
which attain $FP(s_1,k)$ and $FP(s_2,k)$, respectively. The matrix
$
\left[ \begin{array}{cc}
A & \bzero \\
\bzero & B
\end{array} \right]
$
represents a functional $k$-PIR code with $s_1+s_2$ information symbols.

\item[(4)] By (1) and (2) we have that $FP(r,2^r) \leq 2(2^r-1)$ and applying
it $t$ times we obtain $FP(rt,2^r) \leq 2t(2^r-1)$.
\end{enumerate}
\end{IEEEproof}
Our first target in this paper is to improve on Theorem~\ref{thm:simple}(4).

\subsection{Our Contribution and Outline}

In the rest of the paper new lower and upper bounds on $FB(s,k)$ and $FP(s,k)$ will be presented.
In Section \ref{sec:FP} a construction of functional $k$-PIR codes with $k$ being
a power of 2 is presented. Proper puncturing of the code obtained by the construction yields functional $k\text{-PIR}$ codes
for arbitrary $k$. In Section~\ref{sec:FPlower} we provide several lower bound on $FP(s,k)$. First, in Section~\ref{sec:counting}
a general asymptotic lower bound using a counting argument is proved. This argument is applied also on specific values of $s$ and $k$ to get
nontrivial lower bounds on $FP(s,k)$. An improved lower bound for $k=3$ and $k=4$ is presented in Section~\ref{sec:tight4}.
This lower bound is in fact tight. A table on the asymptotic and specific lower and upper bounds for $FP(s,k)$ is
also given. A random construction of functional batch codes is given in Section~\ref{sec:randomFB}.
Bounds on the length of functional batch codes are given in this section too.
In Section~\ref{sec:simplex}, we study the performance of simplex codes when used as functional batch codes.
Conclusions and problems for future research are outlined in Section~\ref{sec:conclude}.

\section{A Construction of Functional PIR Codes}
\label{sec:FP}

In this section an explicit construction of functional $k$-PIR codes when $k$ is a power of 2,
is presented. The code which has $rt$ information symbols
will be represented by two $(t+1) \times 2^r$ arrays.
One array will be defined in the construction and the second array will be defined in the proof for the correctness of the construction.
In the first array, each entry, except for
the entries of the last column, represents the content of different servers.
The last column of the array contains \emph{zeroes}.
In the second array, each column
represents a recovery set. The second array is obtained from the first array by a permutation
defined via a translation induced from the requested symbol.
By puncturing $p$ times this code of length $2^r$, a functional $k$-PIR codes for $k=2^r-2p$ will be obtained.

\begin{construction}
\label{con:PIR2r}
Let $\{x^i_j:1\le i \le t, ~1\le j \le r\}$ be the set of $s=rt$ information symbols.
Let $\cT$ be a $(t+1) \times 2^r$ array whose last column consists of \emph{zeroes}.
The columns of $\cT$ are indexed by the elements of the power set $2^{[r]}$.
The $i$-th row, $1 \leq i \leq t$, contains the $2^r$ linear combinations of the symbols $\{ x^i_j : 1 \leq j \leq r \}$.
In particular, the entry on the column indexed by $A \in 2^{[r]}$ contains the linear combination $x^i_A = \sum_{j\in A} x^i_j$
(note that $x^i_{\varnothing} = 0$). Finally, the $(t+1)$-th row
is a parity row, where the entry in the column indexed by $A$ is
$X_A=\sum_{i=1}^t x^i_A = \sum_{i=1}^t\sum_{j\in A} x^i_j$. This entry will be called the \emph{leader} of the column.
Note that only the entries of the column indexed by $\varnothing$ do not correspond to information stored in a server.
The parity of this column which is \emph{zero} is stored in the $(t+1)$-th row and it is also called a leader.
Each other symbol in the array $\cT$ is stored in a different server. The array $\cT$ contains all the $n=(2^r-1)(t+1)$ symbols and hence it will be called \emph{the stored symbols array}.
\end{construction}

By Theorem~\ref{thm:simple}, $FP(rt,2^r) \leq 2t(2^r-1)$. In the next theorem this upper bound
is improved.

\begin{theorem}
\label{thm:PIR2r}
The code of Construction~\ref{con:PIR2r} is a functional $2^r$-PIR code. Therefore, $FP(rt,2^r)\le (2^r-1)(t+1)$.
\end{theorem}

\begin{IEEEproof}
Let $\mathbf{v}$ be the requested symbol, i.e., $\mathbf{v}$ is a linear combination
$$\mathbf{v}=\mathbf{v}^1+ \mathbf{v}^2  +\cdots+\mathbf{v}^t~,$$
where each $\mathbf{v}^i$ is a linear combination of the information symbols $\{x^i_{j}:1\le j \le r\}$, $1\le i \le t$.
We also define $\bv^{t+1} =0$.

Given the $(t+1) \times 2^r$ stored symbols array $\cT$, we construct a new $(t+1) \times 2^r$ array $\cR^\mathbf{v}$ as follows.
The rows and the columns of $\cR^\mathbf{v}$ are indexed exactly in the same way as the rows and columns of $\cT$ are indexed.
To the symbol in $\cT$ in the entry on the $i$-th row, $1 \leq i \leq t+1$, and the column
indexed by any subset $A$ of $2^{[r]}$, we add $\mathbf{v}^i$ to obtain the corresponding symbol in $\cR^\mathbf{v}$ in the same entry.
The array $\cR^\mathbf{v}$ will be called \emph{the recovery array for $\mathbf{v}$}
since each column contains the content of the servers which form one of the recovery sets.
Note, that the $i$-th row of $\cR^\mathbf{v}$, $1 \leq i \leq t+1$, is a permutation of the $i$-th row of $\cT$ and hence the symbols contained in
$\cR^\mathbf{v}$ are exactly the same symbols contained in $\cT$, which implies that the information of each server is
contained in exactly one entry of $\cR^\mathbf{v}$, but usually not in the same entry as in $\cT$.
The exceptions are the $(t+1)$-th row and each row $i$ for which $\mathbf{v}^i =0$.
It implies that the array $\cR^\mathbf{v}$ represents the content of the servers, but in different entries from those of $\cT$.
We claim now that in each column of $\cR^\mathbf{v}$ contain the content of a set of servers which form a recovery set.

Hence, to complete the proof it is sufficient to show that the sum of the symbols in each column of $\cR^\mathbf{v}$ is $\mathbf{v}$.
For a subset $A$ of $[r]$ let $\cT_A$ be the column of $\cT$ indexed by $A$ and let $\cR^\mathbf{v}_A$ be the column
of $\cR^\mathbf{v}$ indexed by $A$. The sum of the symbols in $\cR^\mathbf{v}_A$ is computed from the symbols
of $\cT_A$ and the request $\mathbf{v}$ as follows
$$\sum_{i=1}^t (x^i_A+\mathbf{v}^i)+X_A= \sum_{i=1}^t x^i_A +X_A + \sum_{i=1}^t \mathbf{v}^i = \sum_{i=1}^t \mathbf{v}^i =\mathbf{v}.$$

Therefore, each column of $\cR^\mathbf{v}$ can serve as a recovery set for the requested symbol $\mathbf{v}$.
Thus, the proof of the theorem is completed.
\end{IEEEproof}

\begin{example}
\label{emp:original}
Let $r=4$, $t=3$, $s=rt=12$, and $k=2^r=16$. All the information symbols
and the coded symbols are represented in the stored symbols array,
where $x^i_{j_1 j_2 \dots j_\ell} \triangleq x^i_{j_1}+x^i_{j_2} + \dots + x^i_{j_\ell}$
and similarly $X_{j_1 j_2 \dots j_\ell} \triangleq \sum_{i=1}^t x^i_{j_1 j_2 \dots j_\ell} =  \sum_{i=1}^t (x^i_{j_1}+x^i_{j_2} + \dots + x^i_{j_\ell})$.

\medskip

\begin{small}
\begin{table}[!h]
\centering
\begin{tabular}{|p{0.65cm}<{\centering}|p{0.65cm}<{\centering}|p{0.65cm}<{\centering}|p{0.65cm}<{\centering}|p{0.65cm}<{\centering}|p{0.65cm}<{\centering}|p{0.65cm}<{\centering}|p{0.65cm}<{\centering}|p{0.65cm}<{\centering}|p{0.65cm}<{\centering}|p{0.65cm}<{\centering}|p{0.65cm}<{\centering}|p{0.65cm}<{\centering}|p{0.65cm}<{\centering}|p{0.65cm}<{\centering}|p{0.65cm}<{\centering}|}
%{|p{0.7cm}|p{0.7cm}|p{0.7cm}|p{0.7cm}|p{0.7cm}|p{0.7cm}|p{0.7cm}|p{0.7cm}|p{0.7cm}|p{0.7cm}|p{0.7cm}|p{0.7cm}|p{0.7cm}|p{0.7cm}|p{0.7cm}|p{0.7cm}|}
\hline
  $x^1_{1}$ & $x^1_{2}$ & $x^1_{3}$ & $x^1_{4}$ & $x^1_{12}$ & $x^1_{13}$ & $x^1_{14}$ & $x^1_{23}$ & $x^1_{24}$ & $x^1_{34}$ & $x^1_{123}$ & $x^1_{124}$ & $x^1_{134}$ & $x^1_{234}$ & $x^1_{1234}$ & 0 \\\hline

  $x^2_{1}$ & $x^2_{2}$ & $x^2_{3}$ & $x^2_{4}$ & $x^2_{12}$ & $x^2_{13}$ & $x^2_{14}$ & $x^2_{23}$ & $x^2_{24}$ & $x^2_{34}$ & $x^2_{123}$ & $x^2_{124}$ & $x^2_{134}$ & $x^2_{234}$ & $x^2_{1234}$ & 0 \\\hline

  $x^3_{1}$ & $x^3_{2}$ & $x^3_{3}$ & $x^3_{4}$ & $x^3_{12}$ & $x^3_{13}$ & $x^3_{14}$ & $x^3_{23}$ & $x^3_{24}$ & $x^3_{34}$ & $x^3_{123}$ & $x^3_{124}$ & $x^3_{134}$ & $x^3_{234}$ & $x^3_{1234}$ & 0 \\\hline

  $X_{1}$ & $X_{2}$ & $X_{3}$ & $X_{4}$ & $X_{12}$ & $X_{13}$ & $X_{14}$ & $X_{23}$ & $X_{24}$ & $X_{34}$ & $X_{123}$ & $X_{124}$ & $X_{134}$ & $X_{234}$ & $X_{1234}$ & 0 \\
  \hline
\end{tabular}
\end{table}
\end{small}

\medskip

Now suppose that the requested symbol is $\mathbf{v}=x^1_1+x^2_{1}+x^2_{2}+x^3_{2}+x^3_{3}+x^3_{4}$, i.e. $\mathbf{v}^1=x^1_1$, $\mathbf{v}^2=x^2_{1}+x^2_{2}$, $\mathbf{v}^3=x^3_{2}+x^3_{3}+x^3_{4}$. For $1\le i \le 3$,
by adding $\mathbf{v}^i$ to each entry in the $i$-th row we obtain the following recovery array.

\medskip

\begin{small}
\begin{table}[!h]
\centering
\begin{tabular}{|p{0.65cm}<{\centering}|p{0.65cm}<{\centering}|p{0.65cm}<{\centering}|p{0.65cm}<{\centering}|p{0.65cm}<{\centering}|p{0.65cm}<{\centering}|p{0.65cm}<{\centering}|p{0.65cm}<{\centering}|p{0.65cm}<{\centering}|p{0.65cm}<{\centering}|p{0.65cm}<{\centering}|p{0.65cm}<{\centering}|p{0.65cm}<{\centering}|p{0.65cm}<{\centering}|p{0.65cm}<{\centering}|p{0.65cm}<{\centering}|}
\hline
  0 & $x^1_{12}$ & $x^1_{13}$ & $x^1_{14}$ & $x^1_{2}$ & $x^1_{3}$ & $x^1_{4}$ & $x^1_{123}$ & $x^1_{124}$ & $x^1_{134}$ & $x^1_{23}$ & $x^1_{24}$ & $x^1_{34}$ & $x^1_{1234}$ & $x^1_{234}$ & $x^1_{1}$ \\\hline

  $x^2_{2}$ & $x^2_{1}$ & $x^2_{123}$ & $x^2_{124}$ & 0 & $x^2_{23}$ & $x^2_{24}$ & $x^2_{13}$ & $x^2_{14}$ & $x^2_{1234}$ & $x^2_{3}$ & $x^2_{4}$ & $x^2_{234}$ & $x^2_{134}$ & $x^2_{34}$ & $x^2_{12}$ \\\hline

  $x^3_{1234}$ & $x^3_{34}$ & $x^3_{24}$ & $x^3_{23}$ & $x^3_{134}$ & $x^3_{124}$ & $x^3_{123}$ & $x^3_{4}$ & $x^3_{3}$ & $x^3_{2}$ & $x^3_{14}$ & $x^3_{13}$ & $x^3_{12}$ & 0 & $x^3_{1}$ & $x^3_{234}$ \\\hline

  $X_{1}$ & $X_{2}$ & $X_{3}$ & $X_{4}$ & $X_{12}$ & $X_{13}$ & $X_{14}$ & $X_{23}$ & $X_{24}$ & $X_{34}$ & $X_{123}$ & $X_{124}$ & $X_{134}$ & $X_{234}$ & $X_{1234}$ & 0 \\
  \hline
\end{tabular}
\end{table}
\end{small}

\medskip

It is straightforward to verify that each column of $\cR^\mathbf{v}$ is a recovery set
for the requested symbol~$\mathbf{v}$. For example, in the third column we have $(x^1_1+x^1_3)+(x^2_1+x^2_2+x^2_3)+(x^3_2+x^3_4)+(x^1_3+x^2_3+x^3_3)=x^1_1+x^2_1+x^2_2+x^3_2+x^3_3+x^3_4=\mathbf{v}$.
\end{example}

%\begin{remark}
%When the number of information symbols satisfies $(t-1)r<s<tr$, we may add some $tr-s$ temporary information symbols and then use the construction above. Then we set those temporary information symbols to be zero and some symbols in the code become zero and thus can be deleted. In this way we have $FP(2t+1,3)\le3t+4$. Combining with the lower bound above we have $FP(2t+1,3)\in\{3t+3,3t+4\}$.
%\end{remark}

\vspace{0.5cm}

The next step is to consider how to modify Construction~\ref{con:PIR2r} for arbitrary $k$.
Since by Theorem~\ref{thm:trivial}(5) $FP(s,2\ell)=FP(s,2\ell-1)+1$ we can consider only even values of $k$.
The main idea is simply to delete some entries of the array $\cT$, i.e. removing some servers
and hence we can say that the $k$-PIR code for $k=2^r$ is being punctured.
This simple idea is less trivial to explain and even less trivial to prove
that the remaining servers can form the required number of recovery sets. Hence,
we start with the simplest case which is $k=2^r-2$ to illustrate the idea.

\begin{construction}
\label{con:1punctured}
Let $\cT$ be the $(t+1) \times 2^r$ stored symbols array constructed in Construction~\ref{con:PIR2r}.
Choose three different subsets $A$, $B$, and $C$ of $[r]$ such that $A = (B \setminus C) \cup (C \setminus B)$.
Delete the symbols in the first $t$ rows of column $\cT_A$ and delete the leader
symbols $X_B$ and $X_C$ in columns $\cT_B$ and $\cT_C$, respectively.
The deletion is done by marking the deleted symbols by a red color. Any deleted
symbol will be also called a \emph{red symbol}. Each deleted symbol is related
to a server which is being removed, i.e.
these $t+2$ red symbols are not associated with any server.
This array obtained from $\cT$ will be denoted by $\tilde{\cT}$ and also called \emph{the stored symbols array}. The
servers store the content of the entries in $\tilde{\cT}$ which are not \emph{zeroes} and do not contain red symbols.
Thus, the length of the code is $n=(t+1)(2^r-1)-(t+2)=(2^r-2)t+2^r-3$.
\end{construction}

\begin{theorem}
\label{thm:1punctured}
The code of Construction~\ref{con:1punctured} is a functional $(2^r-2)$-PIR code.
Therefore, $FP(rt,2^r-2) \le (2^{r}-2)t+2^r-3$.
\end{theorem}

\begin{IEEEproof}
Let $\mathbf{v}$ be the requested symbol, i.e., $\mathbf{v}$ is a linear combination
$$\mathbf{v}=\mathbf{v}^1+\mathbf{v}^2+\dots+\mathbf{v}^t~,$$
where each $\mathbf{v}^i$ is a linear combination of the information symbols $\{x^i_{j}:1\le j \le r\}$, $1\le i \le t$.
We also define $\bv^{t+1} =0$.

Given the $(t+1) \times 2^r$ stored symbols array $\tilde{\cT}$, we construct
a new $(t+1) \times 2^r$ array $\tilde{\cR}^\mathbf{v}$ from $\tilde{\cT}$ exactly
as how $\cR^\mathbf{v}$ was constructed from $\cT$ in the proof of Theorem~\ref{thm:PIR2r}
(adding $\mathbf{v}^i$ to all the $2^r$ entries of the $i$-th row, $1 \leq i \leq t+1$).
The array $\tilde{\cR}^\mathbf{v}$ will be called \emph{the recovery array for $\mathbf{v}$}
since each column without a deleted leader will be used to define a recovery set.
In $\tilde{\cR}^\mathbf{v}$ each symbol in a column of a deleted leader will be called a \emph{free symbol}
since it is free to join any recovery set.
Each symbol which was a red symbol in $\tilde{\cT}$ will maintain
a red symbol in $\tilde{\cR}^\mathbf{v}$ (usually in a different entry, unless
it is either a leader or in the $i$-th row and $\mathbf{v}^i =0$).

Each column with a (non-deleted) leader corresponds to a recovery set as follows.
\begin{itemize}
\item If the column contains no red symbol then the sum of the entries in the column is $\mathbf{v}$
exactly as was proved in Theorem~\ref{thm:PIR2r}.

\item If the column contains a red symbol in
the $i$-th row then we add the symbols of the $i$-th row in columns $\cT_B$ and $\cT_C$ to the
recovery set. The red symbol in the $i$-th row is $x^i_A$. The free symbols in the $i$-th row of columns
$\cT_B$ and $\cT_C$ are $x^i_B+\mathbf{v}^i$ and $x^i_C+\mathbf{v}^i$, respectively.
$x^i_A = x^i_B+ x^i_C = x^i_B+\mathbf{v}^i + x^i_C+\mathbf{v}^i$ and hence the red symbol in the $i$-th row
can be replaced by the related free symbols in columns $\cT_B$ and $\cT_C$. The rest of the proof is as in the proof
of Theorem~\ref{thm:PIR2r}.
\end{itemize}

Therefore, each column of $\cR^\mathbf{v}$ with a (non-deleted) leader can serve as a recovery set for the requested symbol $\mathbf{v}$,
with replaced symbols for possible red symbols in the recovery set.
Thus, the proof of the theorem is completed.
\end{IEEEproof}

\begin{example}
\label{emp:1punctured}
Continuing Example \ref{emp:original} above, choose three subsets $A=\{1234\}$, $B=\{12\}$, and $C=\{34\}$.
Delete the symbols in the first $t$ rows of the column $\cT_A=\cT_{1234}$ and delete the leader symbols
$X_B = X_{12}$ and $X_C = X_{34}$ in columns $\cT_B = \cT_{12}$ and $\cT_C = \cT_{34}$, respectively. The deletion is done by marking the deleted symbols
in a red color. The result is the following stored symbols array.

\medskip

\begin{small}
\begin{table}[!h]
\centering
\begin{tabular}{|p{0.65cm}<{\centering}|p{0.65cm}<{\centering}|p{0.65cm}<{\centering}|p{0.65cm}<{\centering}|p{0.65cm}<{\centering}|p{0.65cm}<{\centering}|p{0.65cm}<{\centering}|p{0.65cm}<{\centering}|p{0.65cm}<{\centering}|p{0.65cm}<{\centering}|p{0.65cm}<{\centering}|p{0.65cm}<{\centering}|p{0.65cm}<{\centering}|p{0.65cm}<{\centering}|p{0.65cm}<{\centering}|p{0.65cm}<{\centering}|}

\hline
  $x^1_{1}$ & $x^1_{2}$ & $x^1_{3}$ & $x^1_{4}$ & $x^1_{12}$ & $x^1_{13}$ & $x^1_{14}$ & $x^1_{23}$ & $x^1_{24}$ & $x^1_{34}$ & $x^1_{123}$ & $x^1_{124}$ & $x^1_{134}$ & $x^1_{234}$ & ${\color{red}x^1_{1234}}$ & 0 \\\hline

  $x^2_{1}$ & $x^2_{2}$ & $x^2_{3}$ & $x^2_{4}$ & $x^2_{12}$ & $x^2_{13}$ & $x^2_{14}$ & $x^2_{23}$ & $x^2_{24}$ & $x^2_{34}$ & $x^2_{123}$ & $x^2_{124}$ & $x^2_{134}$ & $x^2_{234}$ & ${\color{red}x^2_{1234}}$ & 0 \\\hline

  $x^3_{1}$ & $x^3_{2}$ & $x^3_{3}$ & $x^3_{4}$ & $x^3_{12}$ & $x^3_{13}$ & $x^3_{14}$ & $x^3_{23}$ & $x^3_{24}$ & $x^3_{34}$ & $x^3_{123}$ & $x^3_{124}$ & $x^3_{134}$ & $x^3_{234}$ & ${\color{red}x^3_{1234}}$ & 0 \\\hline

  $X_{1}$ & $X_{2}$ & $X_{3}$ & $X_{4}$ & ${\color{red}X_{12}}$ & $X_{13}$ & $X_{14}$ & $X_{23}$ & $X_{24}$ & ${\color{red}X_{34}}$ & $X_{123}$ & $X_{124}$ & $X_{134}$ & $X_{234}$ & $X_{1234}$ & 0 \\
  \hline
\end{tabular}
\end{table}
\end{small}

\medskip

Suppose that the requested symbol is $\mathbf{v}=x^1_1+x^2_{1}+x^2_{2}+x^3_{2}+x^3_{3}+x^3_{4}$,
i.e., $\mathbf{v}^1=x^1_1$, $\mathbf{v}^2=x^2_{1}+x^2_{2}$, $\mathbf{v}^3=x^3_{2}+x^3_{3}+x^3_{4}$.
By adding $\mathbf{v}^i$, $1\le i \le 3$, to each entry in the $i$-th row the following recovery array is obtained.
Note that in this array the deleted symbols are still marked in red, i.e., the red color is with the symbol itself rather than the entry.
Moreover the entries in columns $\cT_B=\cT_{12}$ and $\cT_C=\cT_{34}$ are marked with a yellow color. Since $X_B=X_{12}$ and $X_C=X_{34}$
are deleted, we do not consider using the related columns $\cT_B=\cT_{12}$ and $\cT_C=\cT_{34}$ as recovery sets. Therefore, the symbols on these yellow entries are free
symbols and can be used when we need to replace certain deleted symbols.

\medskip\emph{}

\begin{small}
\begin{table}[!h]
\centering
\begin{tabular}
{|p{0.65cm}<{\centering}|p{0.65cm}<{\centering}|p{0.65cm}<{\centering}|p{0.65cm}<{\centering}|p{0.65cm}<{\centering}|p{0.65cm}<{\centering}|p{0.65cm}<{\centering}|p{0.65cm}<{\centering}|p{0.65cm}<{\centering}|p{0.65cm}<{\centering}|p{0.65cm}<{\centering}|p{0.65cm}<{\centering}|p{0.65cm}<{\centering}|p{0.65cm}<{\centering}|p{0.65cm}<{\centering}|p{0.65cm}<{\centering}|}
\hline
  0 & $x^1_{12}$ & $x^1_{13}$ & $x^1_{14}$ & {\cellcolor{yellow}$x^1_{2}$} & $x^1_{3}$ & $x^1_{4}$ & $x^1_{123}$ & $x^1_{124}$ & {\cellcolor{yellow}$x^1_{134}$} & $x^1_{23}$ & $x^1_{24}$ & $x^1_{34}$ & ${\color{red}x^1_{1234}}$ & $x^1_{234}$ & $x^1_{1}$ \\\hline

  $x^2_{2}$ & $x^2_{1}$ & $x^2_{123}$ & $x^2_{124}$ & {\cellcolor{yellow}0} & $x^2_{23}$ & $x^2_{24}$ & $x^2_{13}$ & $x^2_{14}$ & {\cellcolor{yellow}${\color{red}x^2_{1234}}$} & $x^2_{3}$ & $x^2_{4}$ & $x^2_{234}$ & $x^2_{134}$ & $x^2_{34}$ & $x^2_{12}$ \\\hline

  ${\color{red}x^3_{1234}}$ & $x^3_{34}$ & $x^3_{24}$ & $x^3_{23}$ & {\cellcolor{yellow}$x^3_{134}$} & $x^3_{124}$ & $x^3_{123}$ & $x^3_{4}$ & $x^3_{3}$ & {\cellcolor{yellow}$x^3_{2}$} & $x^3_{14}$ & $x^3_{13}$ & $x^3_{12}$ & 0 & $x^3_{1}$ & $x^3_{234}$ \\\hline

  $X_{1}$ & $X_{2}$ & $X_{3}$ & $X_{4}$ & {\cellcolor{yellow}${\color{red}X_{12}}$} & $X_{13}$ & $X_{14}$ & $X_{23}$ & $X_{24}$ & {\cellcolor{yellow}${\color{red}X_{34}}$} & $X_{123}$ & $X_{124}$ & $X_{134}$ & $X_{234}$ & $X_{1234}$ & 0 \\
  \hline
\end{tabular}
\end{table}
\end{small}

\medskip

As for the deleted (red) symbols located on recovery sets,
the free symbols (symbols in entries marked with yellow) are used to replace the deleted (red) symbols.
For $x^1_{1234}$ and $x^3_{1234}$, the two free symbols in the same row can be used to
replace the deleted (red) symbol, i.e., $x^1_{1234}=x^1_{2}+x^1_{134}$ and $x^3_{1234}=x^3_{134}+x^3_{2}$.
On the second row, the deleted (red) symbol $x^2_{1234}$ lies in an entry marked with yellow and
does not have to be replaced since this column is not used as a recovery set.
Hence, the recovery array is adjusted into the following form. It is then straightforward to verify that
the symbols on each column with an undeleted leader sum up to the requested symbol $\mathbf{v}$.
Therefore, a functional $14$-PIR code is obtained.

\medskip

\begin{table}[!h]
\centering
\begin{tabular}{|p{0.65cm}<{\centering}|p{0.65cm}<{\centering}|p{0.65cm}<{\centering}|p{0.65cm}<{\centering}|p{0.65cm}<{\centering}|p{0.65cm}<{\centering}|p{0.65cm}<{\centering}|p{0.65cm}<{\centering}|p{0.65cm}<{\centering}|p{0.65cm}<{\centering}|p{0.65cm}<{\centering}|p{0.65cm}<{\centering}|p{0.65cm}<{\centering}|p{0.65cm}<{\centering}|p{0.65cm}<{\centering}|p{0.65cm}<{\centering}|}
\hline
  \scalebox{1}{0} & \scalebox{1}{$x^1_{12}$} & \scalebox{1}{$x^1_{13}$} & \scalebox{1}{$x^1_{14}$} & {\cellcolor{yellow}~} & \scalebox{1}{$x^1_{3}$} & \scalebox{1}{$x^1_{4}$} & \scalebox{1}{$x^1_{123}$} & \scalebox{1}{$x^1_{124}$} & {\cellcolor{yellow}~} & \scalebox{1}{$x^1_{23}$} & \scalebox{1}{$x^1_{24}$} & \scalebox{1}{$x^1_{34}$} &\makebox[2.50em][r]{\scalebox{0.9}{$x^1_{2},x^1_{134}$}} & \scalebox{1}{$x^1_{234}$} & \scalebox{1}{$x^1_{1}$}  \\\hline

  \scalebox{1}{$x^2_{2}$} & \scalebox{1}{$x^2_{1}$} & \scalebox{1}{$x^2_{123}$} & \scalebox{1}{$x^2_{124}$} & {\cellcolor{yellow}\scalebox{1}{0}} & \scalebox{1}{$x^2_{23}$} & \scalebox{1}{$x^2_{24}$} & \scalebox{1}{$x^2_{13}$} & \scalebox{1}{$x^2_{14}$} & {\cellcolor{yellow}\scalebox{1}{${\color{red}x^2_{1234}}$}} & \scalebox{1}{$x^2_{3}$} & \scalebox{1}{$x^2_{4}$} & \scalebox{1}{$x^2_{234}$} & \scalebox{1}{$x^2_{134}$} & \scalebox{1}{$x^2_{34}$} & \scalebox{1}{$x^2_{12}$} \\\hline

  \makebox[2.50em][r]{\scalebox{.9}{$x^3_{134},x^3_{2}$}} & \scalebox{1}{$x^3_{34}$} & \scalebox{1}{$x^3_{24}$} & \scalebox{1}{$x^3_{23}$} & {\cellcolor{yellow}~} & \scalebox{1}{$x^3_{124}$} & \scalebox{1}{$x^3_{123}$} & \scalebox{1}{$x^3_{4}$} & \scalebox{1}{$x^3_{3}$} & {\cellcolor{yellow}~} & \scalebox{1}{$x^3_{14}$} & \scalebox{1}{$x^3_{13}$} & \scalebox{1}{$x^3_{12}$} & \scalebox{1}{0} & \scalebox{1}{$x^3_{1}$} & \scalebox{1}{$x^3_{234}$} \\\hline

  \scalebox{1}{$X_{1}$} & \scalebox{1}{$X_{2}$} & \scalebox{1}{$X_{3}$} & \scalebox{1}{$X_{4}$} & {\cellcolor{yellow}\scalebox{1}{${\color{red}X_{12}}$}} & \scalebox{1}{$X_{13}$} & \scalebox{1}{$X_{14}$} & \scalebox{1}{$X_{23}$} & \scalebox{1}{$X_{24}$} & {\cellcolor{yellow}\scalebox{1}{${\color{red}X_{34}}$}}  & \scalebox{1}{$X_{123}$} & \scalebox{1}{$X_{124}$} & \scalebox{1}{$X_{134}$} & \scalebox{1}{$X_{234}$} & \scalebox{1}{$X_{1234}$} & \scalebox{1}{0} \\
  \hline

\end{tabular}
\end{table}

\end{example}

To sum up, the construction of the functional $(2^r-2)$-PIR code is a `1-puncturing' of the functional $(2^r)$-PIR code,
where the punctured symbols are determined by a choice of the tuple of subsets $\{A,B,C\}$.
To generalize this idea to a `p-puncturing', it seems natural to just take more tuples of subsets $\{A_j,B_j,C_j\}$
and perform similar puncturing methods. However, this generalization is non-trivial since one may meet the following scenario.

Say we continue Example \ref{emp:1punctured} and intend to do a `2-puncturing' to obtain a functional $12$-PIR code.
Choose another triple of subsets $\big\{\{13\},\{4\},\{134\}\big\}$. Delete the symbols in the first $t$ rows of the column $\cT_{134}$ and
delete the leader symbols $X_{13}$ and $X_{4}$ in columns $\cT_{13}$ and $\cT_{4}$, respectively.
In the recovering array for the same requested symbol $\mathbf{v}=x^1_1+x^2_{1}+x^2_{2}+x^3_{2}+x^3_{3}+x^3_{4}$, the deleted symbols are marked in red.
The entries in the columns indexed by $\{12\}$, $\{34\}$, $\{13\}$, $\{4\}$ are marked with yellow,
indicating that the symbols on these yellow entries are free symbols and can be used to replace certain deleted symbols.
The recovery array is presented in the following table.

\medskip

\begin{small}
\begin{table}[!h]
\centering
\begin{tabular}{|p{0.65cm}<{\centering}|p{0.65cm}<{\centering}|p{0.65cm}<{\centering}|p{0.65cm}<{\centering}|p{0.65cm}<{\centering}|p{0.65cm}<{\centering}|p{0.65cm}<{\centering}|p{0.65cm}<{\centering}|p{0.65cm}<{\centering}|p{0.65cm}<{\centering}|p{0.65cm}<{\centering}|p{0.65cm}<{\centering}|p{0.65cm}<{\centering}|p{0.65cm}<{\centering}|p{0.65cm}<{\centering}|p{0.65cm}<{\centering}|}
\hline
  0 & $x^1_{12}$ & $x^1_{13}$ & {\cellcolor{yellow}$x^1_{14}$} & {\cellcolor{yellow}$x^1_{2}$} & {\cellcolor{yellow}$x^1_{3}$} & $x^1_{4}$ & $x^1_{123}$ & $x^1_{124}$ & {\cellcolor{yellow}${\color{red}x^1_{134}}$} & $x^1_{23}$ & $x^1_{24}$ & $x^1_{34}$ & ${\color{red}x^1_{1234}}$ & $x^1_{234}$ & $x^1_{1}$ \\\hline

  $x^2_{2}$ & $x^2_{1}$ & $x^2_{123}$ & {\cellcolor{yellow}$x^2_{124}$} & {\cellcolor{yellow}0} & {\cellcolor{yellow}$x^2_{23}$} & $x^2_{24}$ & $x^2_{13}$ & $x^2_{14}$ & {\cellcolor{yellow}${\color{red}x^2_{1234}}$} & $x^2_{3}$ & $x^2_{4}$ & $x^2_{234}$ & ${\color{red}x^2_{134}}$ & $x^2_{34}$ & $x^2_{12}$ \\\hline

  ${\color{red}x^3_{1234}}$ & $x^3_{34}$ & $x^3_{24}$ & {\cellcolor{yellow}$x^3_{23}$} & {\cellcolor{yellow}${\color{red}x^3_{134}}$} & {\cellcolor{yellow}$x^3_{124}$} & $x^3_{123}$ & $x^3_{4}$ & $x^3_{3}$ & {\cellcolor{yellow}$x^3_{2}$} & $x^3_{14}$ & $x^3_{13}$ & $x^3_{12}$ & 0 & $x^3_{1}$ & $x^3_{234}$ \\\hline

  $X_{1}$ & $X_{2}$ & $X_{3}$ & {\cellcolor{yellow}${\color{red}X_{4}}$} & {\cellcolor{yellow}${\color{red}X_{12}}$} & {\cellcolor{yellow}${\color{red}X_{13}}$} & $X_{14}$ & $X_{23}$ & $X_{24}$ & {\cellcolor{yellow}${\color{red}X_{34}}$} & $X_{123}$ & $X_{124}$ & $X_{134}$ & $X_{234}$ & $X_{1234}$ & 0 \\
  \hline
\end{tabular}
\end{table}
\end{small}

\medskip

Now, on each row there are two deleted (red) symbols that should be replaced by
combinations of free symbols in yellow entries. The problem is that we cannot simply replace
$x^1_{1234}$ with $x^1_{2}+x^1_{134}$ as before in Example \ref{emp:1punctured} since now
$x^1_{134}$ is also a deleted symbol. The solution is to replace $x^1_{1234}$ by $x^1_{2}+x^1_{14}+x^1_{3}$
and $x^1_{134}$ does not need repairing since it lies on a yellow entry.
This scenario demonstrates that generalizing `1-puncturing' into `p-puncturing' is nontrivial
in the sense that we need an explicit algorithm to describe how to use the free symbols to replace the deleted symbols.

Our generalization of Construction~\ref{con:1punctured} and the proof of its correctness
in Theorem~\ref{thm:1punctured}, i.e. generalizing the 1-puncturing to $p$-puncturing,
will consist of four steps. In the first step, $p$ pairwise disjoint triples from $2^{[r]}$ will be defined
(two elements of a triple for deleting two leader symbols and the third one for deleting the symbols of the column excluding the leader).
In the second step the related recovery array is constructed similarly to the definition in Construction~\ref{con:1punctured}.
In the third step a replacing operation (in several rounds) to replace the deleted (red) symbols by free symbols will be described. In the last step we will prove that these
actual replacements result in the required recovery sets.

Following these ideas, Construction~\ref{con:1punctured} for $k$-PIR, $k=2^r-2$ can be generalized to arbitrary $k=2^r-2p$,
where $1 \leq p < 2^{r-2}$. For the first step of the construction (defining the pairwise disjoint triples)
we need the following definition and results on \emph{partial spreads}.

\begin{definition}
A partial $2$-spread of $\F_2^r$ is a collection of $2$-dimensional subspaces $V_1,\dots,V_M$ of $\F_2^r$ such that $V_i \cap V_j=\{\mathbf{0}\}$ for all $i\neq j$.
\end{definition}

It is shown in~\cite{EtVa11} that a partial 2-spread with $M = 2^{r-2}$ always exists.
In each $2$-dimensional subspace $V_i$ we have three nonzero vectors.
Let $A_i$, $B_i$ and $C_i$ be their supports which are subsets of $[r]$. By the definition of a partial $2$-spread,
$A_i = (B_i \setminus C_i) \cup (C_i \setminus B_i  )$ and the triples $\{\{A_i,B_i,C_i\}:1\le i \le M\}$ are pairwise disjoint.

\begin{construction}
\label{con:p-punctured}
Let $\cT$ be the $(t+1) \times 2^r$ stored symbols array constructed in Constructions~\ref{con:PIR2r} and~\ref{con:1punctured}.
Since $p < 2^{r-2}$ there exists a partial 2-spread $\F_2^r$ which contains
$p$ pairwise disjoint triples $\{\{A_i,B_i,C_i\}:1\le i \le p\}$
such that $A_i = (B_i \setminus C_i ) \cup (C_i \setminus B_i)$.
For each triple $\{A_i,B_i,C_i\}$, delete the symbols in the first $t$ rows of column $\cT_{A_i}$ and
the leader symbols $X_{B_i}$ and $X_{C_i}$ in columns $\cT_{B_i}$ and $\cT_{C_i}$, respectively.
The deletion is done by marking the deleted symbols by a red color. Any deleted
symbol will be called a \emph{red symbol},
These $p(t+2)$ red symbols are not associated with any server.
This array obtained from $\cT$ will be denoted by $\tilde{\cT}$.
Thus, the length of the code is $n=(t+1)(2^r-1)-tp-2p=(2^r-p-1)t+2^r-2p-1$.
\end{construction}

\begin{theorem}
\label{thm:p-punctured}
The code of Construction~\ref{con:p-punctured} is a functional $(2^r-2p)$-PIR code. Therefore, $FP(rt,2^r-2p) \le (2^r-p-1)t+2^r-2p-1$, for $0\le p <2^{r-2}$.
\end{theorem}

\begin{IEEEproof}
Let $\mathbf{v}$ be the requested symbol, i.e., $\mathbf{v}$ is a linear combination
$$\mathbf{v}=\mathbf{v}^1+\dots+\mathbf{v}^t~,$$
where each $\mathbf{v}^i$ is a linear combination of the information symbols $\{x^i_{j}:1\le j \le r\}$.
We also define $\bv^{t+1} =0$.

Given the $(t+1) \times 2^r$ stored symbols array $\tilde{\cT}$, we construct a new $(t+1) \times 2^r$ array $\tilde{\cR}^\mathbf{v}$ exactly
as in the proofs of Theorems~\ref{thm:PIR2r} and~\ref{thm:1punctured} (adding $\mathbf{v}^i$ to all the $2^r$ entries of the $i$-th row, $1 \leq i \leq t+1$).
Each symbol which was a red symbol in $\tilde{\cT}$ will be also a red symbol in $\tilde{\cR}^\mathbf{v}$ (usually in a different entry, unless
it is either a leader or in the $i$-th row and $\mathbf{v}^i =0$).

The $2^r -2p$ recovery sets relate to the $2^r - 2p$ columns in which the leaders were not deleted.
By the proof of Theorem~\ref{thm:PIR2r}, the sum of the symbols (including the red ones) in each such column is $\bv$.
Our goal is that each column whose leader was not deleted will be a recovery set. Hence, we have to
apply a procedure to replace the red symbols in these columns.
For each row $i$, $1 \leq i \leq t$, we apply the following procedure. In each step of the procedure the number of red
symbols in the row will be the same as the number of pairs of columns with deleted leaders
which have some symbols (red or free). Before the first step the number of red symbols in the row is $p$
and the number of such pairs is also $p$.

Let $\{B_j,C_j\}$ be a pair from the disjoint triples for which the two related columns do not contain a red symbol.
If there is no such pair then all the red symbols are in the columns with deleted leaders
and the procedure for the row is completed.
The sum of the symbols in column $B_j$ (of the $i$-th row) is $x^i_{B_j}+\mathbf{v}^i$ and in
column $C_j$ is $x^i_{B_j}+\mathbf{v}^i$.
$x^i_{B_j}+\mathbf{v}^i +x^i_{B_j}+\mathbf{v}^i = x^i_{A_j}$ and $x^i_{A_j}$ is a red symbol
in some column $D$ (neither $B_j$ nor $C_j$ (since the related two entries do not have a red symbol).
We replace the red symbol $x^i_{A_j}$ of column $D$ with the two symbols $x^i_{B_j}+\bv^i$
and $x^i_{C_j}+\bv^i$ (which are not marked in red). Entries $B_j$ and $C_j$ in the $i$-th row will become empty.
The number of red symbols in the $i$-th row was reduced by one and also the number of pairs of columns
with deleted leader which have some symbols was reduced by one. Hence, these number remain equal and this property
is satisfied at the end of the step for this row.
Note, that the red symbol $x^i_{A_j}$ was replaced by two free (non-red) symbols whose sum equal to $x^i_{A_j}$.

After this procedure was applied on all the first $t$ rows, all the recovery sets will not contain any red symbols.
The sum of symbols of any recovery set is not changed during the procedure.
The non-red symbols in new constructed array $\hat{\cR}^\bv$ are the same as the non-red symbols in $\tilde{\cR}^\mathbf{v}$.

Therefore, each column of $\hat{\cR}^\bv$ can serve as a recovery set for the requested symbol $\bv$.
Thus, the proof of the theorem is completed.
\end{IEEEproof}

\begin{example}
Continuing Example \ref{emp:1punctured}, choose three disjoint triples of subsets $\{\{12\}, \{34\}, \{1234\}\}$, $\{\{13\},\{4\},\{134\}\}$ and $\{\{2\},\{3\},\{23\}\}$.
Delete the symbols in the first $t$ rows of the columns $\cT_{1234}$, $\cT_{134}$ and $\cT_{23}$.
Delete the leader symbols $X_{12}$, $X_{34}$, $X_{13}$, $X_{4}$, $X_{2}$ and $X_{3}$. The deletion is done by marking the deleted symbols
in a red color. The result is the following stored symbols array.

\medskip

\begin{small}
\begin{table}[!h]
\centering
\begin{tabular}{|p{0.65cm}<{\centering}|p{0.65cm}<{\centering}|p{0.65cm}<{\centering}|p{0.65cm}<{\centering}|p{0.65cm}<{\centering}|p{0.65cm}<{\centering}|p{0.65cm}<{\centering}|p{0.65cm}<{\centering}|p{0.65cm}<{\centering}|p{0.65cm}<{\centering}|p{0.65cm}<{\centering}|p{0.65cm}<{\centering}|p{0.65cm}<{\centering}|p{0.65cm}<{\centering}|p{0.65cm}<{\centering}|p{0.65cm}<{\centering}|}
\hline
  $x^1_{1}$ & $x^1_{2}$ & $x^1_{3}$ & $x^1_{4}$ & $x^1_{12}$ & $x^1_{13}$ & $x^1_{14}$ & ${\color{red}x^1_{23}}$ & $x^1_{24}$ & $x^1_{34}$ & $x^1_{123}$ & $x^1_{124}$ & ${\color{red}x^1_{134}}$ & $x^1_{234}$ & ${\color{red}x^1_{1234}}$ & 0 \\\hline

  $x^2_{1}$ & $x^2_{2}$ & $x^2_{3}$ & $x^2_{4}$ & $x^2_{12}$ & $x^2_{13}$ & $x^2_{14}$ & ${\color{red}x^2_{23}}$ & $x^2_{24}$ & $x^2_{34}$ & $x^2_{123}$ & $x^2_{124}$ & ${\color{red}x^2_{134}}$ & $x^2_{234}$ & ${\color{red}x^2_{1234}}$ & 0 \\\hline

  $x^3_{1}$ & $x^3_{2}$ & $x^3_{3}$ & $x^3_{4}$ & $x^3_{12}$ & $x^3_{13}$ & $x^3_{14}$ & ${\color{red}x^3_{23}}$ & $x^3_{24}$ & $x^3_{34}$ & $x^3_{123}$ & $x^3_{124}$ & ${\color{red}x^3_{134}}$ & $x^3_{234}$ & ${\color{red}x^3_{1234}}$ & 0 \\\hline

  $X_{1}$ & ${\color{red}X_{2}}$ & ${\color{red}X_{3}}$ & ${\color{red}X_{4}}$ & ${\color{red}X_{12}}$ & ${\color{red}X_{13}}$ & $X_{14}$ & $X_{23}$ & $X_{24}$ & ${\color{red}X_{34}}$ & $X_{123}$ & $X_{124}$ & $X_{134}$ & $X_{234}$ & $X_{1234}$ & 0 \\
  \hline
\end{tabular}
\end{table}
\end{small}

\medskip

Suppose that the requested symbol is $\mathbf{v}=x^1_1+x^2_{1}+x^2_{2}+x^3_{2}+x^3_{3}+x^3_{4}$,
i.e., $\mathbf{v}^1=x^1_1$, $\mathbf{v}^2=x^2_{1}+x^2_{2}$, $\mathbf{v}^3=x^3_{2}+x^3_{3}+x^3_{4}$.
By adding $\mathbf{v}^i$, $1\le i \le 3$, to each entry in the $i$-th row the following recovery array is obtained.
Note that in this array the deleted symbols are still marked in red.
Moreover the entries in columns $\cT_{12}$, $\cT_{34}$, $\cT_{13}$, $\cT_{4}$, $\cT_{2}$ and $\cT_{3}$ are marked with a yellow color.
Since $X_{12}$, $X_{34}$, $X_{13}$, $X_{4}$, $X_{2}$ and $X_{3}$ are deleted,
we do not consider using the related columns as recovery sets. Therefore, the symbols on yellow entries are free
symbols and can be used when we need to replace certain deleted symbols.

\medskip

\begin{small}
\begin{table}[!h]
\centering
\begin{tabular}{|p{0.65cm}<{\centering}|p{0.65cm}<{\centering}|p{0.65cm}<{\centering}|p{0.65cm}<{\centering}|p{0.65cm}<{\centering}|p{0.65cm}<{\centering}|p{0.65cm}<{\centering}|p{0.65cm}<{\centering}|p{0.65cm}<{\centering}|p{0.65cm}<{\centering}|p{0.65cm}<{\centering}|p{0.65cm}<{\centering}|p{0.65cm}<{\centering}|p{0.65cm}<{\centering}|p{0.65cm}<{\centering}|p{0.65cm}<{\centering}|}
\hline
  0 & {\cellcolor{yellow}$x^1_{12}$} & {\cellcolor{yellow}$x^1_{13}$} & {\cellcolor{yellow}$x^1_{14}$} & {\cellcolor{yellow}$x^1_{2}$} & {\cellcolor{yellow}$x^1_{3}$} & $x^1_{4}$ & $x^1_{123}$ & $x^1_{124}$ & {\cellcolor{yellow}${\color{red}x^1_{134}}$} & ${\color{red}x^1_{23}}$ & $x^1_{24}$ & $x^1_{34}$ & ${\color{red}x^1_{1234}}$ & $x^1_{234}$ & $x^1_{1}$ \\\hline

  $x^2_{2}$ & {\cellcolor{yellow}$x^2_{1}$} & {\cellcolor{yellow}$x^2_{123}$} & {\cellcolor{yellow}$x^2_{124}$} & {\cellcolor{yellow}0} & {\cellcolor{yellow}${\color{red}x^2_{23}}$} & $x^2_{24}$ & $x^2_{13}$ & $x^2_{14}$ & {\cellcolor{yellow}${\color{red}x^2_{1234}}$} & $x^2_{3}$ & $x^2_{4}$ & $x^2_{234}$ & ${\color{red}x^2_{134}}$ & $x^2_{34}$ & $x^2_{12}$ \\\hline

  ${\color{red}x^3_{1234}}$ & {\cellcolor{yellow}$x^3_{34}$} & {\cellcolor{yellow}$x^3_{24}$} & {\cellcolor{yellow}${\color{red}x^3_{23}}$} & {\cellcolor{yellow}${\color{red}x^3_{134}}$} & {\cellcolor{yellow}$x^3_{124}$} & $x^3_{123}$ & $x^3_{4}$ & $x^3_{3}$ & {\cellcolor{yellow}$x^3_{2}$} & $x^3_{14}$ & $x^3_{13}$ & $x^3_{12}$ & 0 & $x^3_{1}$ & $x^3_{234}$ \\\hline

  $X_{1}$ & {\cellcolor{yellow}${\color{red}X_{2}}$} & {\cellcolor{yellow}${\color{red}X_{3}}$} & {\cellcolor{yellow}${\color{red}X_{4}}$} & {\cellcolor{yellow}${\color{red}X_{12}}$} & {\cellcolor{yellow}${\color{red}X_{13}}$} & $X_{14}$ & $X_{23}$ & $X_{24}$ & {\cellcolor{yellow}${\color{red}X_{34}}$} & $X_{123}$ & $X_{124}$ & $X_{134}$ & $X_{234}$ & $X_{1234}$ & 0 \\
  \hline
\end{tabular}
\end{table}
\end{small}

\medskip

Independently, on each row red symbols are replaced step by step, e.g., the third row is transformed step by step as follows:

\medskip

\begin{small}
\begin{table}[!h]
\centering
\begin{tabular}{|p{0.65cm}<{\centering}|p{0.65cm}<{\centering}|p{0.65cm}<{\centering}|p{0.65cm}<{\centering}|p{0.65cm}<{\centering}|p{0.65cm}<{\centering}|p{0.65cm}<{\centering}|p{0.65cm}<{\centering}|p{0.65cm}<{\centering}|p{0.65cm}<{\centering}|p{0.65cm}<{\centering}|p{0.65cm}<{\centering}|p{0.65cm}<{\centering}|p{0.65cm}<{\centering}|p{0.65cm}<{\centering}|p{0.65cm}<{\centering}|}
\hline

  ${\color{red}x^3_{1234}}$ & {\cellcolor{yellow}$x^3_{34}$} & {\cellcolor{yellow}$x^3_{24}$} & {\cellcolor{yellow}${\color{red}x^3_{23}}$} & {\cellcolor{yellow}${\color{red}x^3_{134}}$} & {\cellcolor{yellow}$x^3_{124}$} & $x^3_{123}$ & $x^3_{4}$ & $x^3_{3}$ & {\cellcolor{yellow}$x^3_{2}$} & $x^3_{14}$ & $x^3_{13}$ & $x^3_{12}$ & 0 & $x^3_{1}$ & $x^3_{234}$ \\\hline

  \multicolumn{16}{c}{$\Downarrow$~$\Downarrow$~$\Downarrow$}\\\hline

  \scalebox{1}{${\color{red}x^3_{1234}}$} & {\cellcolor{yellow}~} & {\cellcolor{yellow}~} & \scalebox{.95}{\cellcolor{yellow}\makebox[2.80em][r]{$x^3_{34},x^3_{24}$}}  & {\cellcolor{yellow}\scalebox{1}{${\color{red}x^3_{134}}$}} & {\cellcolor{yellow}\scalebox{1}{$x^3_{124}$}} & \scalebox{1}{$x^3_{123}$} & \scalebox{1}{$x^3_{4}$} & \scalebox{1}{$x^3_{3}$} & {\cellcolor{yellow}\scalebox{1}{$x^3_{2}$}} & \scalebox{1}{$x^3_{14}$} & \scalebox{1}{$x^3_{13}$} & \scalebox{1}{$x^3_{12}$} & \scalebox{1}{0} & \scalebox{1}{$x^3_{1}$} & \scalebox{1}{$x^3_{234}$} \\\hline

  \multicolumn{16}{c}{$\Downarrow$~$\Downarrow$~$\Downarrow$}\\\hline

  \scalebox{1}{${\color{red}x^3_{1234}}$} & {\cellcolor{yellow}~} & {\cellcolor{yellow}~} & {\cellcolor{yellow}~} &
   \makebox[2.8em][r]{\scalebox{.9}{ {\cellcolor{yellow} \tabincell{c} {$x^3_{34},x^3_{24}$ \\ $x^3_{124}$}} }}

  & {\cellcolor{yellow}~} & \scalebox{1}{$x^3_{123}$} & \scalebox{1}{$x^3_{4}$} & \scalebox{1}{$x^3_{3}$} & {\cellcolor{yellow}\scalebox{1}{$x^3_{2}$}} & \scalebox{1}{$x^3_{14}$} & \scalebox{1}{$x^3_{13}$} & \scalebox{1}{$x^3_{12}$} & \scalebox{1}{0} & \scalebox{1}{$x^3_{1}$} & \scalebox{1}{$x^3_{234}$} \\\hline

  \multicolumn{16}{c}{$\Downarrow$~$\Downarrow$~$\Downarrow$}\\\hline

  %\makebox[2.9em][r]{\scalebox{.49}{$x^3_{34},x^3_{24},x^3_{124},x^3_2$}}

   \makebox[2.8em][r]{\scalebox{.9}{ \tabincell{c} {$x^3_{34},x^3_{24}$ \\ $x^3_{124},x^3_2$} }} & {\cellcolor{yellow}~} & {\cellcolor{yellow}~} & {\cellcolor{yellow}~} & {\cellcolor{yellow}~} & {\cellcolor{yellow}~} & \scalebox{1}{$x^3_{123}$} & \scalebox{1}{$x^3_{4}$} & \scalebox{1}{$x^3_{3}$} & {\cellcolor{yellow}~} & \scalebox{1}{$x^3_{14}$} & \scalebox{1}{$x^3_{13}$} & \scalebox{1}{$x^3_{12}$} & \scalebox{1}{0} & \scalebox{1}{$x^3_{1}$} & \scalebox{1}{$x^3_{234}$} \\\hline
\end{tabular}

\end{table}
\end{small}

\medskip

After the appropriate red symbols were replaced in all the rows, the recovery array is as follows.

\medskip

\begin{table}[!h]
\centering

\begin{small}
\begin{tabular}{|p{0.65cm}<{\centering}|p{0.65cm}<{\centering}|p{0.65cm}<{\centering}|p{0.65cm}<{\centering}|p{0.65cm}<{\centering}|p{0.65cm}<{\centering}|p{0.65cm}<{\centering}|p{0.65cm}<{\centering}|p{0.65cm}<{\centering}|p{0.65cm}<{\centering}|p{0.65cm}<{\centering}|p{0.65cm}<{\centering}|p{0.65cm}<{\centering}|p{0.65cm}<{\centering}|p{0.65cm}<{\centering}|p{0.65cm}<{\centering}|}
\hline

  0 & {\cellcolor{yellow}~} & {\cellcolor{yellow}~} & {\cellcolor{yellow}~} & {\cellcolor{yellow}~} & {\cellcolor{yellow}~} & $x^1_{4}$ & $x^1_{123}$ & $x^1_{124}$ & {\cellcolor{yellow}~} &
  \makebox[2.5em][r]{\scalebox{.9}{$x^1_{12},x^1_{13}$}} & $x^1_{24}$ & $x^1_{34}$ & \makebox[2.6em][r]{\scalebox{.9}{ \tabincell{c} {$x^1_{2},x^1_{14}$ \\ $x^1_{3}$} }} & $x^1_{234}$ & $x^1_{1}$ \\\hline

  $x^2_{2}$ & {\cellcolor{yellow}~} & {\cellcolor{yellow}~} & {\cellcolor{yellow}~} & {\cellcolor{yellow}0} & {\cellcolor{yellow}~} & ${x^2_{24}}$ & $x^2_{13}$ & $x^2_{14}$ & {\cellcolor{yellow}${\color{red}x^2_{1234}}$} & $x^2_{3}$ & $x^2_{4}$ & $x^2_{234}$ & \makebox[2.7em][r]{\scalebox{.9}{ \tabincell{c} {$x^2_{124},x^2_{1}$ \\ $x^2_{123}$} }} & $x^2_{34}$ & $x^2_{12}$ \\\hline

  \makebox[2.8em][r]{\scalebox{.9}{ \tabincell{c} {$x^3_{34},x^3_{24}$ \\ $x^3_{124},x^3_2$} }} & {\cellcolor{yellow}~} & {\cellcolor{yellow}~} & {\cellcolor{yellow}~} & {\cellcolor{yellow}~} & {\cellcolor{yellow}~} & \scalebox{1}{$x^3_{123}$} & \scalebox{1}{$x^3_{4}$} & \scalebox{1}{$x^3_{3}$} & {\cellcolor{yellow}~} & \scalebox{1}{$x^3_{14}$} & \scalebox{1}{$x^3_{13}$} & \scalebox{1}{$x^3_{12}$} & \scalebox{1}{0} & \scalebox{1}{$x^3_{1}$} & \scalebox{1}{$x^3_{234}$} \\\hline

  $X_{1}$ & {\cellcolor{yellow}${\color{red}X_{2}}$} & {\cellcolor{yellow}${\color{red}X_{3}}$} & {\cellcolor{yellow}${\color{red}X_{4}}$} & {\cellcolor{yellow}${\color{red}X_{12}}$} & {\cellcolor{yellow}${\color{red}X_{13}}$} & $X_{14}$ & $X_{23}$ & $X_{24}$ & {\cellcolor{yellow}${\color{red}X_{34}}$} & $X_{123}$ & $X_{124}$ & $X_{134}$ & $X_{234}$ & $X_{1234}$ & 0 \\
  \hline
\end{tabular}
\end{small}
\end{table}

It is straightforward to verify that the symbols on each column with undeleted leader sum up to the requested symbol $\mathbf{v}$.
\end{example}

\medskip

As mentioned in Theorem \ref{thm:trivial}, by deleting any symbol in a functional $(2^r-2p)$-PIR code we obtain a functional $(2^r-2p-1)$-PIR code, therefore we have
\begin{corollary}
$FP(rt,2^r-2p-1)\le(2^{r}-1-p)t+2^r-2p-2$, for $0\le p <2^{r-2}$.
\end{corollary}

\begin{remark}
\label{rem:tune}
Note, that all the constructions above for functional $k$-PIR codes with ${k\in[2^{r-1}+1,2^r]}$ are described for $rt$ information symbols.
When the number of information symbols is not a multiple of $r$, say $rt+r'$, $0<r'<r$, we may add $r-r'$ virtual information
symbols and apply the constructions above. All the virtual information symbols are set to zero. Also some coded symbols, which are linear
combinations of only virtual information symbols are set to zero.

For example, assume we want to construct a functional $2^r$-PIR code of dimension $rt+r'$.
We add $r-r'$ virtual information symbols and hence we construct a functional $2^r$-PIR code of dimension $r(t+1)$ of length $(2^r-1)(t+2)$ using Construction~\ref{con:PIR2r}.
The virtual information symbols are now set to zero and thus some $c$ symbols (linear combinations of virtual information symbols) are set to zero.
The number $c$ is $2^{r-r'}-1$ when $t\ge1$ or $2^{r-r'+1}-2$ when $t=0$ (since some `leader' symbols are also set to zero when $t=0$).
Therefore for any $0<r'<r$, $FP(r',2^r)\le 2(2^r-2^{r-r'})$ and $FP(rt+r',2^r)\le(2^r-1)(t+1)+2^r-2^{r-r'}$ when $t\ge 1$.
\end{remark}

Similar idea holds when $2^{r-1}<k<2^r$, but this should be done carefully, since the $c$ symbols set to zero are dependent on
the way that the puncturing from the functional $2^r$-PIR code to the functional $k$-PIR code is done.
Following this way, some results with small parameters are summarized in Table~\ref{tab:tab1}.

\begin{table}[!h]
\centering
\caption{Upper bounds on $FP(s,k)$ arising from our construction in Section~\ref{sec:FP}.}
  \begin{tabular}{|c|c|c|}
     \hline
     ~ & $k=6$ & $k=8$ \\\hline
     $s=1$ & $6$ & $8$ \\\hline
     $s=2$ & $9$ & $12$ \\\hline
     $s=3t~(t\ge1)$ & $6t+5$ & $7t+7$ \\\hline
     $s=3t+1~(t\ge1)$ & $6t+8$ & $7t+11$ \\\hline
     $s=3t+2~(t\ge1)$ & $6t+10$ & $7t+13$ \\\hline
   \end{tabular}~
   \begin{tabular}{|c|c|c|c|c|}
     \hline
     ~ & $k=10$ & $k=12$ & $k=14$ & $k=16$ \\\hline
     $s=1$ & $10$ & $12$ & $14$ & $16$ \\\hline
     $s=2$ & $15$ & $18$ & $21$ & $24$ \\\hline
     $s=3$ & $19$ & $22$ & $25$ & $28$ \\\hline
     $s=4t~(t\ge1)$ & $12t+9$ & $13t+11$ & $14t+13$ & $15t+15$ \\\hline
     $s=4t+1~(t\ge1)$ & $12t+14$ & $13t+17$ & $14t+20$ & $15t+23$ \\\hline
     $s=4t+2~(t\ge1)$ & $12t+18$ & $13t+21$ & $14t+24$ & $15t+27$ \\\hline
     $s=4t+3~(t\ge1)$ & $12t+20$ & $13t+23$ & $14t+26$ & $15t+29$ \\\hline
   \end{tabular}
\smallskip
\label{tab:tab1}
\end{table}

\section{Lower Bounds on the Length of Functional PIR Codes}
\label{sec:FPlower}

This section is devoted to lower bounds on the length of functional PIR codes.
When the number of requests $k$ is a fixed constant\footnote{more precisely $k=o(s)$.}, $P(s,k)=s+o(s)$ (see Lemma \ref{lem:known})
and hence the research objective is to
analyze the redundancy part $o(s)$. However, for functional
PIR codes this is not the case. By using a counting argument it will be proved in this section that $FP(s,k)$ grows
linearly in $s$, i.e., $\lim_{s\rightarrow\infty} FP(s,k)/s \geq c$ for some constant $c$ to be determined. Using another approach in this section,
a better lower bound on $FP(s,3)$ and $FP(s,4)$ is derived. Codes for $k=4$ in Construction~\ref{con:PIR2r} attain this bound and hence
the bound is exact.

\subsection{A general lower bound by counting}
\label{sec:counting}

In our exposition which follows we will need some properties of the binomial coefficients.
These properties are proved in the following lemmas.

\begin{lemma}
\label{lem:twice_bin}
If $n > 3r+2$ then $\binom{n}{r+1} > 2 \binom{n}{r}$.
\end{lemma}
\begin{IEEEproof}
Follows immediately by comparing $\binom{n}{r+1}$ with $2 \binom{n}{r}$.
\end{IEEEproof}

\begin{lemma}
\label{lem:bin_sum}
If $n > 3r+2$ then $\binom{n}{r+1} > \sum_{i=1}^r \binom{n}{i}$.
\end{lemma}
\begin{IEEEproof}
By induction on $r$, where the basis is $\binom{n}{2} > \binom{n}{1}$ and in the induction step
Lemma~\ref{lem:twice_bin} is used.
\end{IEEEproof}

\begin{lemma}
\label{lem:lem_binom}
If $n > 3r+2$ then $\binom{n}{r+1} > \sum_{i=1}^{r-1} (r-i) \binom{n}{i}$.
\end{lemma}
\begin{IEEEproof}
Again, by induction on $r$, where the basis for $r=2$, where $\binom{n}{3} > \binom{n}{1}$.
For the induction hypothesis assume that the claim is true for $r-1$, i.e.
$$\binom{n}{r} > \sum_{i=1}^{r-2} (r-1-i) \binom{n}{i}~.$$
By Lemma~\ref{lem:bin_sum} we have
$$\binom{n}{r+1} > \sum_{i=1}^r \binom{n}{i}~,$$
and combining this with the induction hypothesis we have
$$\binom{n}{r+1} > \sum_{i=1}^r \binom{n}{i} > \sum_{i=1}^{r-2} (r-1-i) \binom{n}{i} + \sum_{i=1}^{r-1} \binom{n}{i}
= \sum_{i=1}^{r-2} (r-i) \binom{n}{i} + \binom{n}{r-1} = \sum_{i=1}^{r-1} (r-i) \binom{n}{i}~,$$
which proves the induction step.
\end{IEEEproof}

For the next theorem we
remind the reader that by Theorem~\ref{thm:trivial}(5) we have
$FP(s,2\ell)=FP(s,2\ell-1)+1$ and hence can consider only even values of $k$.
The even values will be considered since they imply better bounds than the
related odd values.

\begin{theorem}
\label{thm:lower_bound_FP}
For a fixed even integer $k\ge 4$,
$$\lim_{s\rightarrow\infty} \frac{FP(s,k)}{s} \ge \frac{1}{H(1/k)}~,$$
where $H(\cdot)$ is the binary entropy function defined by $H(p)=-p\log{p}-(1-p)\log{(1-p)}$.
\end{theorem}

\begin{IEEEproof}
Suppose there exists a functional $k$-PIR code of dimension $s$ and length $n$.
For each request $\bv$, we have $k$ disjoint recovery sets of $[n]$. The sum of the sizes of
all these $k (2^s-1)$ recovery sets is at most $n(2^s-1)$.
Hence, the average size of a recovery set should be at most~$\frac{n}{k}$.

Consider all the subsets of $[n]$ of size at most $\lceil \frac{n}{k} \rceil +1$. If each such subset is used as a recovery
set for some request, then the average size of a recovery set is at least
\begin{equation}
\label{eq:av_rec}
\frac{\sum_{i=1}^{\lceil \frac{n}{k} \rceil +1} i\binom{n}{i}}{\sum_{i=1}^{\lceil \frac{n}{k} \rceil +1} \binom{n}{i}}
\end{equation}
By applying Lemma~\ref{lem:lem_binom} on $\binom{n}{\lceil \frac{n}{k} \rceil+1}$ we have
\begin{equation}
\label{eq:eq_binom}
\binom{n}{\lceil \frac{n}{k} \rceil+1} > \sum_{i=1}^{\lceil \frac{n}{k} \rceil-1} (\left\lceil \frac{n}{k} \right\rceil-i) \binom{n}{i}
\end{equation}
By developing the numerator in (\ref{eq:av_rec}) and plugging (\ref{eq:eq_binom}) in the process we obtain
$$
\sum_{i=1}^{\lceil \frac{n}{k} \rceil +1} i\binom{n}{i}
= \sum_{i=1}^{\lceil \frac{n}{k} \rceil} i\binom{n}{i}
+ \left\lceil \frac{n}{k} \right\rceil  \binom{n}{\lceil \frac{n}{k} \rceil+1} + \binom{n}{\lceil \frac{n}{k} \rceil+1}
$$
$$
> \sum_{i=1}^{\lceil \frac{n}{k} \rceil} i\binom{n}{i} + \left\lceil \frac{n}{k} \right\rceil  \binom{n}{\lceil \frac{n}{k} \rceil+1}
+ \sum_{i=1}^{\lceil \frac{n}{k} \rceil-1} (\left\lceil \frac{n}{k} \right\rceil-i) \binom{n}{i} = \left\lceil \frac{n}{k} \right\rceil
\sum_{i=1}^{\lceil \frac{n}{k} \rceil +1} \binom{n}{i}~.
$$
Now, we can evaluate the average in (\ref{eq:av_rec}) as
$$
\frac{\sum_{i=1}^{\lceil \frac{n}{k} \rceil +1} i\binom{n}{i}}{\sum_{i=1}^{\lceil \frac{n}{k} \rceil +1} \binom{n}{i}}
>  \frac{\left\lceil \frac{n}{k} \right\rceil \sum_{i=1}^{\lceil \frac{n}{k} \rceil +1} \binom{n}{i}}{\sum_{i=1}^{\lceil \frac{n}{k} \rceil +1} \binom{n}{i}}
=\left\lceil \frac{n}{k} \right\rceil \geq \frac{n}{k},
$$
which contradicts our proof that the average size of a recovery set is at most $\frac{n}{k}$.

Therefore, not all the subsets of $[n]$ of size at most $\lceil \frac{n}{k} \rceil +1$ are
used as recovery sets,
which implies that $\sum_{i=1}^{\lceil \frac{n}{k} \rceil+1} \binom{n}{i}>k(2^s-1)$.
The left hand side tends to $2^{n H(1/k)}$ as $n$ tends to infinity. Hence, if $n=cs$, then
$$
2^{cs H(1/k)} >k(2^s-1)~,
$$
which implies that $c H(1/k)>1$ and the claim of the theorem follows.
\end{IEEEproof}

Note, that the counting argument used in the proof of Theorem~\ref{thm:lower_bound_FP} implies
that the recovery sets used for all the possible requests are of the smallest possible size.
In practice, it is difficult to assume that this would be the case. Improving the lower bound
by taking larger recovery sets into account is a future task.

The first several lower bounds on $\lim_{s\rightarrow\infty} \frac{FP(s,k)}{s}$ derived
from Theorem~\ref{thm:lower_bound_FP}, together with the related
upper bounds implied by Construction \ref{con:p-punctured}, are
summarized in Table~\ref{table:FP}. The lower bound for $k=4$ will be further improved in Section~\ref{sec:tight4}.

\begin{center}
\begin{table}[!h]
\caption{Lower and upper bounds on $\lim_{s\rightarrow\infty} \frac{FP(s,k)}{s}$ }\label{table:FP}
\begin{tabular}{|c|p{1.4cm}<{\centering}|p{1.4cm}<{\centering}|p{1.4cm}<{\centering}|p{1.4cm}<{\centering}|p{1.4cm}<{\centering}|p{1.4cm}<{\centering}|p{1.4cm}<{\centering}|p{1.4cm}<{\centering}|}
\hline
  $k$  & 2 & 4 & 6 & 8 & 10 & 12 & 14 & 16 \\\hline
  \text{lower~bound} & 1 & 1.2326 & 1.5384 & 1.8397 & 2.1322 & 2.4165 & 2.6937 & 2.9648 \\\hline
  \text{upper~bound} & 1 & 1.5 & 2 & 2.3333 & 3 & 3.25 & 3.5 & 3.75 \\\hline
  $k$ & 18 & 20 & 22 & 24 & 26 & 28 & 30 & 32\\\hline
  \text{lower~bound} & 3.2306 & 3.4917 & 3.7486 & 4.0019 & 4.2518 & 4.4987 & 4.7429 & 4.9845\\\hline
  \text{upper~bound} & 4.8 & 5 & 5.2 & 5.4 & 5.6 & 5.8 & 6 & 6.2 \\\hline
\end{tabular}
\smallskip
\end{table}
\end{center}

The technique used in the proof of Theorem~\ref{thm:lower_bound_FP} can be applied
slightly differently to obtain lower bounds on $FP(s,k)$
for specific parameters $s$ and $k$.

Suppose we have a functional $k$-PIR code with dimension $s$ and length $n$.
For each request $\bv$, we have $k$ disjoint subsets of $[n]$,
$R_1,\dots,R_k$, where each one of them is a recovery set for $\bv$.
For each such request $\bv$ we choose arbitrarily such $k$ recovery sets.
Therefore, $k(2^s-1)$ distinct recovery sets are chosen.
Let $\Lambda (s)$ be the sum of the size of all these recovery sets.
Since the $k$ recovery sets $R_1,\dots,R_k$ for any request $\bv$ are pairwise disjoint, it follows that
$$\sum_{i=1}^k | R_i| \leq n~,$$
which implies that
\begin{equation}
\label{eq:sumR}
\Lambda (s) \leq n(2^s-1)~.
\end{equation}
On the other hand, a lower bound of $\Lambda (s)$ can be obtained by choosing the recovery sets with smallest size as possible,
since the size of the recovery sets by such a choice will be a lower bound on the actual size.
There are $k(2^s-1)$ distinct recovery sets. Let $d$ be the largest integer such that
\begin{equation}
\label{eq:sum_for_k}
\sum_{i=1}^{d} {n\choose i}\le k(2^s-1)~.
\end{equation}
The smallest lower bound $\Lambda (s)$ will be obtained if all the
$\sum_{i=1}^{d} \binom{n}{i}$ subsets of size $d$ or less will be included as
recovery sets. It implies that in the chosen $k(2^s-1)$ recovery sets, at least
$k(2^s-1)-\sum_{i=1}^{d} \binom{n}{i}$ subsets of size $d+1$ or greater than $d+1$, are included
to obtain the lower bound.
Therefore,

\begin{equation}
\label{eq:counting}
\sum_{i=1}^{d} i \binom{n}{i}+(d+1)\bigg(k(2^s-1)-\sum_{i=1}^{d} \binom{n}{i} \bigg) \le \Lambda (s).
\end{equation}

The lower bound on $FP(s,k)$ is obtained by comparing (\ref{eq:sumR}) and (\ref{eq:counting}), i.e.,
finding the minimum~$n$ for which
$$
\sum_{i=1}^{d} i \binom{n}{i}+(d+1)\bigg(k(2^s-1)-\sum_{i=1}^{d} \binom{n}{i} \bigg) \leq n(2^s-1).
$$

%For example, we have $FP(9,6)\leq 23$ by Construction ???. Assume that $n=20$,
%and apply~(\ref{eq:counting}) for $s=9$ and $k=6$ and $n=20$. We have that
%$$\binom{20}{1} + \binom{20}{2} + \binom{20}{3} = 1350 < 6 \cdot (2^9-1) = 3066$$
%and
%$$\binom{20}{1} + \binom{20}{2} + \binom{20}{3} + \binom{20}{4} = 6195 > 6 \cdot (2^9-1) = 3066~.$$
%Since in a code of length 20 has $6 \cdot (2^9-1) = 3066$ recovery sets, it follows that
%there are at least $3066 - 1350 = 1716$ recovery sets of size at least four.
%Therefore by~(\ref{eq:counting}),
%$$\binom{20}{1} + 2 \binom{20}{2} + 3 \binom{20}{3} + 4 \cdot 1716 = 10684 \leq \Lambda (6) ~,$$
%which is a contradiction to $\Lambda (6) \le 20 \cdot (2^9 -1)=10220$ by~(\ref{eq:sumR}).
%Thus, $FP(9,6) > 20$.

\begin{example}
\label{ex:FP6_8}
Assume that $FP(6,8)=20$,
and apply~(\ref{eq:sum_for_k}) for $s=6$, $k=8$ and $n=20$, i.e.,
$$\binom{20}{1} + \binom{20}{2} = 210 < 8 \cdot (2^6-1) = 504$$
and
$$
\binom{20}{1} + \binom{20}{2} + \binom{20}{3} = 1350 > 8 \cdot (2^6-1) = 504~.
$$
Since in this code of length 20, a total of $8 \cdot (2^6-1) = 504$ recovery sets are required, it follows that
there are at least $504 - 210 = 294$ recovery sets of size at least three.
Therefore by~(\ref{eq:counting}),
$$\binom{20}{1} + 2 \binom{20}{2} + 3 \cdot 294 = 1282 \leq \Lambda (6) ~,$$
which is a contradiction to $\Lambda (6) \le 20 \cdot (2^6 -1)=1260$ by~(\ref{eq:sumR}).
Thus, $FP(6,8) > 20$ and since by Theorem \ref{thm:PIR2r}, $FP(6,8)\leq 21$, it follows that $FP(6,8)=21$.
\end{example}

%\vspace{0.1cm}

\begin{example}
\label{ex:FP2_k}
When $k$ is even we have $FP(2,k) \leq \frac{3k}{2}$ (encode the two information
symbols $x_1$ and $x_2$ into $x_1$, $x_2$, and $x_1+x_2$; each one of these three encoded symbol will appear $\frac{k}{2}$ times in the code.)

Assume now that $n = FP(2,k) \leq \frac{3k}{2}-1$ and apply~(\ref{eq:counting}) for $s=2$, $k$ and $n=\frac{3k}{2}-1$.
For each request, three recovery sets are required for a total of $3k$ recovery sets. There are at most $n=\frac{3k}{2}-1$ recovery sets of size 1.
Therefore, there are at least $\frac{3k}{2}+1$ recovery sets whose size at least two.
Hence, by~(\ref{eq:counting}),
$$
(\frac{3k}{2}-1) + 2 \cdot (\frac{3k}{2}+1) = 3 \cdot \frac{3k}{2} +1 \leq \Lambda (2) ~.
$$
By~(\ref{eq:sumR}), $\Lambda (2) \le n \cdot (2^2 -1)= 3 \cdot \frac{3k}{2} -3$, a contradiction.

Therefore, $FP(2,k) > \frac{3k}{2}-1$ and thus $FP(2,k)=\frac{3k}{2}$ when $k$ is even.
\end{example}

Table~\ref{table:num} contains some specific bounds on $FP(s,k)$ for $s\le 32$ and $6\le k\le 16$, where $k$ is even.

\begin{table}[!h]
\caption{Numerical results on $FP(s,k)$} \label{table:num}
\begin{tabular}{|c|p{2cm}<{\centering}|p{2cm}<{\centering}|p{2cm}<{\centering}|p{2cm}<{\centering}|p{2cm}<{\centering}|p{2cm}<{\centering}|}
  \hline
  \diagbox{$s$}{$k$} & 6 & 8 & 10 & 12 & 14 & 16 \\\hline
  1 & 6 & 8 & 10 & 12 & 14 & 16 \\\hline
  2 & 9 & 12 & 15 & 18 & 21 & 24 \\\hline
  3 & 11 & 14 & 18-19 & 21-22 & 25 & 28 \\\hline
  4 & 12-14 & 15-18 & 19-21 & 23-24 & 27 & 30 \\\hline
  5 & 15-16 & 18-20 & 22-26 & 25-30 & 28-34 & 31-38 \\\hline
  6 & 16-17 & 21 & 25-30 & 29-34 & 33-38 & 37-42 \\\hline
  7 & 17-20 & 22-25 & 27-32 & 32-36 & 37-40 & 41-44 \\\hline
  8 & 19-22 & 23-27 & 29-33 & 34-37 & 39-41 & 44-45 \\\hline
  9 & 21-23 & 26-28 & 31-38 & 35-43 & 41-48 & 46-53 \\\hline
  10 & 22-26 & 28-32 & 34-42 & 39-47 & 43-52 & 47-57 \\\hline
  11 & 24-28 & 30-34 & 36-44 & 42-49 & 47-54 & 52-59 \\\hline
  12 & 26-29 & 31-35 & 38-45 & 45-50 & 51-55 & 57-60 \\\hline
  13 & 28-32 & 34-39 & 39-50 & 46-56 & 53-62 & 60-68 \\\hline
  14 & 29-34 & 36-41 & 42-54 & 47-60 & 55-66 & 62-72 \\\hline
  15 & 30-35 & 38-42 & 45-56 & 51-62 & 57-68 & 63-74 \\\hline
  16 & 32-38 & 39-46 & 47-57 & 55-63 & 61-69 & 67-75 \\\hline
  17 & 34-40 & 41-48 & 49-62 & 57-69 & 65-76 & 72-83 \\\hline
  18 & 35-41 & 44-49 & 50-66 & 59-73 & 67-80 & 75-87 \\\hline
  19 & 37-44 & 46-53 & 54-68 & 60-75 & 69-82 & 78-89 \\\hline
  20 & 39-46 & 47-55 & 56-69 & 64-76 & 71-83 & 79-90 \\\hline
  21 & 40-47 & 49-56 & 58-74 & 67-82 & 75-90 & 82-98 \\\hline
  22 & 41-50 & 51-60 & 59-78 & 69-86 & 79-94 & 87-102 \\\hline
  23 & 43-52 & 53-62 & 62-80 & 71-88 & 81-96 & 91-104 \\\hline
  24 & 45-53 & 55-63 & 65-81 & 73-89 & 83-97 & 93-105 \\\hline
  25 & 46-56 & 56-67 & 67-86 & 77-95 & 84-104 & 95-113 \\\hline
  26 & 47-58 & 59-69 & 69-90 & 80-99 & 89-108 & 97-117 \\\hline
  27 & 49-59 & 61-70 & 70-92 & 82-101 & 92-110 & 102-119 \\\hline
  28 & 51-62 & 62-74 & 73-93 & 83-102 & 95-111 & 106-120 \\\hline
  29 & 52-64 & 63-76 & 76-98 & 85-108 & 97-118 & 108-128 \\\hline
  30 & 54-65 & 66-77 & 78-102 & 89-112 & 98-122 & 110-132 \\\hline
  31 & 56-68 & 68-81 & 79-104 & 92-114 & 103-124 & 111-134 \\\hline
  32 & 57-70 & 70-83 & 81-105 & 94-115 & 106-125 & 116-135 \\\hline
\end{tabular}
\smallskip

The exact values for $s=1$ are trivial and the exact values for $s=2$ are given in Example~\ref{ex:FP2_k}.
For $s\ge 3$, the lower bounds are derived by the counting method while the upper bounds are by the main construction in
Theorem~\ref{thm:p-punctured} and Remark~\ref{rem:tune}.
\end{table}

\subsection{A tight bound of $FP(s,3)$ and $FP(s,4)$}
\label{sec:tight4}

This subsection is devoted to analyzing $FP(s,3)$ and $FP(s,4)$. Recall that by Lemma~\ref{lem:systematic},
a functional PIR code can be always assumed to be systematic.

Let ${t\brace b}$ be the Stirling number of the second kind, which calculates the number
of partitions of $[t]$ into $b$ nonempty subsets. It is well known that
$$ {t\brace b}=\frac{1}{b!}\sum_{i=0}^b (-1)^{b-i}{b\choose i} i^t. $$

Now, we derive the following lower bound on $FP(s,3)$.

\begin{theorem}
\label{thm:FP3}
For any given $s \geq 3$ we have that $FP(s,3)\ge
  \begin{cases}
    \frac{3}{2}s+2 & \mbox{if s is even}  \\
    \frac{3}{2}(s+1) & \mbox{if s is odd}
  \end{cases}~.$
\end{theorem}

\begin{IEEEproof}
Clearly, $FP(s,3)=s+t$, where $t \geq 0$. The $s \times (s+t)$ matrix $\bG$ representing the functional $3$-PIR code is of the
form $\bG=[\bI_s~\bU ]$, where $\bI_s$ is the $s \times s$ identity matrix.
The columns of $\bU$ are denoted by $\{\bu_1,\dots,\bu_t\}$.

A nonzero requested (column) vector $\bv$ can be recovered as
$\bv=\sum_{i\in I_1} \be_i+\sum_{j\in U_1} \bu_j=\sum_{i\in I_2} \be_i+\sum_{j\in U_2} \bu_j=\sum_{i\in I_3} \be_i+\sum_{j\in U_3} \bu_j$,
where $I_1,I_2,I_3$ are three pairwise disjoint subsets of $[s]$ and $U_1,U_2,U_3$ are three pairwise disjoint subsets of $[t]$.
The unordered triple $\{U_1,U_2,U_3\}$ will be called a \emph{feasible triple} corresponding to the requested vector $\bv$.
W.l.o.g. if we have $U_1=U_2=\varnothing$ then $I_1$ and $I_2$ have the same indices for unit vectors
which sum to $\bv$, contradicting the disjointness of $I_1$ and $I_2$.
Therefore, in a feasible triple at most one of $U_1,U_2,U_3$ is empty.

Next, it is claimed that no two requested vectors share a common feasible triple.

To prove the claim let $\{U_1,U_2,U_3\}$ be a feasible triple and let $\bw_j$ be the sum of the columns related to $U_j$, $1 \leq j \leq 3$.
The requested vector $\bv$ is recovered based on $\bw_1$, $\bw_2$ and $\bw_3$ and some unit vectors.
Note that each $\be_i$ can be used only once to recover $\bv$. Therefore, $\bw_1$, $\bw_2$ and $\bw_3$ determine a unique request vector $\bv$.
This can be observed as follows by considering each coordinate of $\bv$ and the related coordinate in $\bw_1$, $\bw_2$, and $\bw_3$.
Consider now the $i$-th coordinate, $1 \leq i \leq s$.

Assume the triple obtained from the value of the triple ($\bw_1$,$\bw_2$,$\bw_3$)
in the $i$-th coordinate is $(0,0,1)$. If the $i$-th coordinate
of $\bv$ is \emph{one} then we must have $\be_i$ in both $I_1$ and $I_2$,
contradicting the fact that $\be_i$ can be used only once. Therefore,
the value of the $i$-th coordinate of $\bv$ is \emph{zero}.

Similarly, the value of the $i$-th coordinate of $\bv$ is \emph{zero} if the value of the triple ($\bw_1$,$\bw_2$,$\bw_3$) in the $i$-th
coordinate is $(0,1,0)$, $(1,0,0)$, or $(0,0,0)$.
The value of the $i$-th coordinate of $\bv$ is \emph{one} if the value of the triple ($\bw_1$,$\bw_2$,$\bw_3$) in the $i$-th
coordinate is $(0,1,1)$, $(1,0,1)$, $(1,1,0)$, or $(1,1,1)$.

Therefore, the requested vector $\bv$ is uniquely determined by $U_1$, $U_2$, and $U_3$. Thus,
no two requested vectors share a common feasible triple which completes the proof of the claim.

Let $U_4 \triangleq [t] \setminus (U_1\bigcup U_2 \bigcup U_3)$ and distinguish between the following four cases
in counting the number of feasible triples $\{U_1,U_2,U_3\}$:

\begin{enumerate}
\item If each one of $U_1$, $U_2$, $U_3$, and $U_4$ is nonempty, then
the number of feasible triples is the same as the number of partitions of $[t]$ into
four nonempty subsets, where one of them is chosen to be $U_4$.
The number of such partitions, i.e. feasible triples, is $4{t\brace 4}$.

\item If each of $U_1$, $U_2$, and $U_3$ is nonempty and $U_4$ is empty, then
the number of feasible triples is the same as the number of partitions of $[t]$ into
three nonempty subsets.
Hence, number of such feasible triples is ${t\brace 3}$.

\item If exactly one of $U_1$, $U_2$, and $U_3$ is empty and $U_4$ is nonempty, then
the number of feasible triples is the same as the number of partitions of $[t]$ into
three nonempty subsets, where one of them is chosen to be $U_4$.
Hence, the number of such feasible triple is $3{t\brace 3}$.

\item If exactly one of $U_1$, $U_2$, and $U_3$ is empty and $U_4$ is empty, then
the number of feasible triples is the same as the number of partitions of $[t]$ into
two nonempty subsets.
Therefore, number of such feasible triples is ${t\brace 2}$.
\end{enumerate}

Thus, the number of feasible triples is at most
$$4{t\brace 4}+4{t\brace 3}+{t\brace 2}=\frac{4^t}{6}-2^{t-1}+\frac{1}{3}.$$
On the other hand, we proved that no two requested vectors share a common feasible triple.
Hence, there are at least $2^s-1$ feasible triples and this implies that
$$2^s-1\le \frac{4^t}{6}-2^{t-1}+\frac{1}{3}~.$$
Thus, $t>\frac{s+\log 6}{2}$.
\end{IEEEproof}

The lower bound of Theorem~\ref{thm:FP3} can be combined with the bounds of
Theorem~\ref{thm:trivial} to obtain lower bounds on $FP(s,k)$ for $k > 3$.
In particular we have.

\begin{corollary}
\label{cor:FP3}
For any $s \geq 3$ we have $FP(s,4)\ge
  \begin{cases}
    \frac{3}{2}s+3 & \mbox{if s is even}  \\
    \frac{3}{2}(s+1)+1 & \mbox{if s is odd}
  \end{cases}~.$
\end{corollary}

Considering Theorem~\ref{thm:FP3}, Theorem~\ref{thm:PIR2r}, Theorem~\ref{thm:trivial}, Corollary~\ref{cor:FP3}
and Remark~\ref{rem:tune}, we have that

\begin{corollary}
For any $t \geq 2$, $FP(2t,3)=3t+2$, $FP(2t,4)=3t+3$, $3t+3\le FP(2t+1,3)\le 3t+4$ and $3t+4\le FP(2t+1,4)\le3t+5$.
\end{corollary}

\section{Bounds on the Length of Functional Batch Codes}
\label{sec:randomFB}

In this section a random construction of functional batch codes is presented.
The random construction relies on a well-known result of random constructions
for linear codes which attain the sphere-covering bound~\cite{Blin87,Blin90}.

\begin{definition}
For a binary code $\cC$ of length $n$, the \emph{covering radius} is the smallest integer $R$ such that
for any $\bv\in\F_2^{n}$, there exists $\bu\in\cC$ such that $d(\bv,\bu)\le R$.
The code $\cC$ is a code with covering radius $R$.
\end{definition}

\begin{proposition}\cite{CKMS85}
If $\cC$ is a binary linear code of length $n$, and dimension $k$,
with a parity check matrix $\bH$, then $\cC$ has covering radius $R$
if and only if every column vector $\F_2^{n-k}$ is the sum of at most $R$ columns of $\bH$.
\end{proposition}

Let $V(n,R)$ be the size of the Hamming ball of radius R. A code
with covering radius $R$ has at least $\frac{2^n}{V(n,R)}$ codewords
and thus a linear code with covering radius $R$ has dimension $k\ge n-\log V(n,R)$. This is the sphere covering bound for linear codes.
Blinovskii~\cite{Blin87,Blin90} proved that almost all linear codes attain the sphere covering bound
(see also~\cite[Ch. 12, p. 325]{CHLLbook} and the references therein).

\begin{theorem}
\label{thm:covering}
Let $0\le \rho < 1/2$, $\cC_{k,n}$ be the ensemble of $2^{kn}$ linear
codes generated by all possible binary $k\times n$ matrices,
and $R_n=\lfloor\rho n\rfloor$. There exists a sequence $k_n$ for which
$$
k_n/n\le1-H(\rho)+O(n^{-1}\log n)~,
$$
such that the fraction of codes $C_n \in \cC_{k_n,n}$
which have covering radius $R_n$ tends to 1, when $n$ tends to infinity.
\end{theorem}

In other words, Theorem~\ref{thm:covering} implies that if a binary random matrix $\cH$ of size $s\times n$
is considered as a parity check matrix of a linear code,
then the covering radius $R=\rho n$ of the code satisfies $H(\rho)\sim\frac{s}{n}$ with probability tending to 1,
when $n$ tends to infinity, i.e., any column vector of length $s$ is the sum of at most $R$ columns of $\cH$.

Cooper~\cite{Cooper00} proved the following result on the invertibility of random binary matrices.

\begin{theorem}
\label{thm:random}
Let $\bG$ be a random binary matrix of size $s\times s$, where each entry is independently and identically
distributed with $\text{Pr}[\bG_{i,j}=1]=p(s)$. If $\min\{p(s),1-p(s)\}\ge (\log s+d(s))/s$ for
any $d(s)\rightarrow\infty$, then $\text{Pr}[\bG\text{~is invertible}]$
tends to a constant $c\approx 0.28879$, when $s$ tends to infinity.
\end{theorem}

We are now in a position to present the random construction of functional batch codes.
The idea is illustrated first with an example on functional 2-batch codes.
For sufficiently large $s$, randomly choose a binary matrix of size $s\times n$ to represent the functional 2-batch code.
Let $\bu,\bv$ be two arbitrary requests. By Theorem \ref{thm:covering},
with probability tending to 1, when $s$ and $n$ tend to infinity, the request $\bu$ can be recovered as a sum
of $\rho n$ columns, where $H(\rho)\sim\frac{s}{n}$. The remaining matrix is a random
matrix of size $s\times (1-\rho)n$. If $(1-\rho)n>s$, then by Theorem~\ref{thm:random},
it has an $s\times s$ invertible sub-matrix with probability $c\approx 0.28879$ .
Using the columns from this invertible sub-matrix, the request $\bv$ can be recovered.
Therefore, under the constraints $(1-\rho)n>s$, $H(\rho)\sim\frac{s}{n}$, there exists
a binary matrix of size $s\times n$ representing a functional 2-batch code when~$s$ and~$n$ are sufficiently large.
To find the asymptotic relation between~$n$ and~$s$, note that the constraints require $s/n\sim H(\rho)<1-\rho$.
The root of $1-\rho=H(\rho)$ is $\rho=0.227$ and thus we can set $n\sim1.2937s$.
The next theorem generalizes this idea to arbitrary functional $k$-batch codes.

\begin{theorem}
\label{theorem:UFB}
If $c_1=\frac{1}{2}$ and $c_{k+1}$ is the root of the polynomial $H(z)=H(c_k)-zH(c_k)$,
then
$$
\lim_{s\rightarrow\infty} \frac{FB(s,k)}{s}\le \frac{1}{H(c_{k})}.
$$
\end{theorem}

\begin{IEEEproof}
For a sufficiently large $s$, randomly choose an $s\times n_1$ binary matrix $\bG_1$ to represent
the functional $k$-batch code. With probability tending to 1 the
first request can be recovered as a sum of $\rho_1 n_1$ columns of $\bG_1$, where $H(\rho_1)\sim\frac{s}{n_1}$.
Let $\bG_2$ be the matrix obtained by removing these $\rho_1 n_1$ columns from $\bG_1$.
$\bG_2$ is an $s\times n_2$ random matrix, where
$n_2=(1-\rho_1)n_1$. The second request can be recovered, with probability which tends to 1,
as a sum of $\rho_2 n_2$ columns on $\bG_2$, where $H(\rho_2) \sim \frac{s}{n_2}$.
This procedure continues and for the $j$-th request, $1\le j \le k-1$, we have a matrix $\bG_j$.
The $j$-th request can be recovered, with probability tending to 1, as a sum of
$\rho_j n_j$ columns, where $H(\rho_j)\sim\frac{s}{n_j}$ and $n_j=\prod_{i=1}^{j-1} (1-\rho_i)n_1$.
Finally, for the $k$-th request, we have to show that the remaining matrix $\bG_k$ contains
an $s \times s$ invertible sub-matrix. This is guaranteed by Theorem~\ref{thm:random} with positive probability
$c\approx 0.28879$ as long as $s<n_k=\prod_{i=1}^{k-1} (1-\rho_i)n_1$ for
sufficiently large $s$. Therefore, we have a binary matrix of size
$s\times n_1$ representing a functional $k$-batch code if $s<n_k=\prod_{i=1}^{k-1} (1-\rho_i)n_1$.

To complete the proof we have to derive the asymptotic relation between $n_1$ and $s$. Note first that
$$
\frac{s}{n_1}\sim H(\rho_1)\sim H(\rho_2)(1-\rho_1)\sim \cdots\sim H(\rho_{k-1})\prod_{i=1}^{k-2} (1-\rho_i)<\prod_{i=1}^{k-1} (1-\rho_i).
$$

Hence, to maximize $\frac{s}{n_1}$, we should have $H(\rho_{k-1})=1-\rho_{k-1}$, $H(\rho_{k-2})=H(\rho_{k-1})(1-\rho_{k-2})$, $\dots$, $H(\rho_j)=H(\rho_{j+1})(1-\rho_j)$, $\dots$, $H(\rho_1)=H(\rho_2)(1-\rho_1)$. Therefore,
we set $\rho_{k-1}=c_2,~\rho_{k-2}=c_3,~\dots,~\rho_{1}=c_k$ and thus asymptotically we have $n_1 \sim \frac{s}{H(c_{k})}$.
\end{IEEEproof}

A lower bound of $FB(s,k)$ can be derived as follows.

\begin{theorem}
\label{theorem:LFB}
$$\lim_{s\rightarrow\infty} \frac{FB(s,k)}{s} \ge \frac{k}{\log(k+1)}.$$
\end{theorem}

\begin{IEEEproof}
Assume there is a functional $k$-batch code of length $n$ and dimension $s$, represented by an $s\times n$ matrix $\bG$.
For any recovery process of a request $\bv = (\bv_1,\dots,\bv_k)$ with $k$ vectors of length $s$, assign a label to each column of $\bG$.
The label is either $0$ or some $i$, $1\le i \le k$. A label 0 indicates that the column is not used in the recovery process of $\bv$.
A label $i$, indicates that the column is used in the recovery set for $\bv_i$. Then the labeling
of $\bG$ for the request $\bv$ is an element in $\{0,1,\dots,k\}^n$. For any two different ordered $k$-tuples of request vectors
$(\bv_1,\dots,\bv_k)$ and $(\bu_1,\dots,\bu_k)$, where $\bv_1,\dots,\bv_k$ are $k$ distinct vectors
and $\bu_1,\dots,\bu_k$ are also $k$ distinct vectors, the labeling of $\bG$ must be different. Therefore,
$(k+1)^n \geq {2^{s}-1 \choose k} k!$.

Thus,
$$\lim_{s\rightarrow\infty} \frac{n}{s} \ge \frac{k}{\log(k+1)},$$
which completes the proof.
\end{IEEEproof}

\vspace{0.3cm}

Table \ref{table:FB} summarizes the lower and upper bounds of $\lim_{s\rightarrow\infty} \frac{FB(s,k)}{s}$.

\begin{table}[!h]
\caption{Lower and upper bounds of $\lim_{s\rightarrow\infty} \frac{FB(s,k)}{s}$ (by Theorems~\ref{theorem:UFB} and~\ref{theorem:LFB})}
\label{table:FB}
\centering
\begin{tabular}{|c|c|c|c|c|c|}
\hline
  $k$ & 2 & 3 & 4 & 5 & 6 \\\hline
  $\lim_{s\rightarrow\infty} \frac{FB(s,k)}{s}$ & 1.2619-1.2937 & 1.5000-1.5489 & 1.7227-1.7828 & 1.9343-2.0028 & 2.1372-2.2124  \\\hline
  $k$ & 7 & 8 & 9 & 10 & 11 \\\hline
  $\lim_{s\rightarrow\infty} \frac{FB(s,k)}{s}$ & 2.3333-2.4137 & 2.5237-2.6089 & 2.7093-2.7984 & 2.8906-2.9834 & 3.0684-3.1641  \\\hline
  $k$ & 12 & 13 & 14 & 15 & 16 \\\hline
  $\lim_{s\rightarrow\infty} \frac{FB(s,k)}{s}$ & 3.2429-3.3414 & 3.4144-3.5156 & 3.5834-3.6869 & 3.7500-3.8557 & 3.9144-4.0222  \\\hline
  $k$ & 17 & 18 & 19 & 20 & 21  \\\hline
  $\lim_{s\rightarrow\infty} \frac{FB(s,k)}{s}$ & 4.0768-4.1865 & 4.2374-4.3489 & 4.3962-4.5094 & 4.5534-4.6683 & 4.7091-4.8256  \\\hline
  $k$ & 22 & 23 & 24 & 25 & 26 \\\hline
  $\lim_{s\rightarrow\infty} \frac{FB(s,k)}{s}$ & 4.8634-4.9814 & 5.0164-5.1358 & 5.1681-5.2889 & 5.3187-5.4407 & 5.4681-5.5914  \\\hline
  $k$ & 27 & 28 & 29 & 30 & 31 \\\hline
  $\lim_{s\rightarrow\infty} \frac{FB(s,k)}{s}$ & 5.6164-5.7410 & 5.7637-5.8895 & 5.9101-6.0369 & 6.0555-6.1835 & 6.2000-6.3291  \\\hline
\end{tabular}
\smallskip
\end{table}

\section{Using Simplex Codes as Functional Batch Codes}\label{sec:simplex}

In~\cite{FVY15a} it was shown that $P(r,2^{r-1}) = 2^r-1$ and in~\cite{WKCB17} it was proved that $B(r,2^{r-1}) = 2^r-1$. Furthermore, in Theorem~\ref{thm:simple}, we also confirmed that $FP(r,2^{r-1}) = 2^r-1$. Hence, in this section we analyze whether the same property is valid also for functional batch codes, that is, whether the property $FB(r,2^{r-1}) = 2^r-1$ holds. These three results were proved using simplex codes, which are defined as follows.
\begin{definition}
A $[2^r-1,r]$ simplex code is a linear code of length $n=2^r-1$ and dimension~$r$ whose $r \times n$ generator matrix $\mathbf{G}$ contains each nonzero column vector $\mathbf{z}$ of length $r$ exactly once as a column.
\end{definition}

Simplex codes have been used for several more applications, among them are write-once memory (WOM) codes and random I/O (RIO) codes. An $[n,k,t]$ WOM code is a coding scheme comprising of $n$ binary cells such that it is possible to write a $k$-bit message $t$ times while on each write the cell values can only change from zero to one. An $(n,k,t)$ RIO code assumes that $t$ $k$-bit messages are stored in $n$ cells each with $t+1$ levels such that every page can be read by sensing a single read threshold. In~\cite{YM16}, it was proved that these two families of codes are equivalent and a new variation of RIO codes, called \emph{parallel RIO codes}, has been proposed, where all messages  can be written together and thereby can allow the design of codes with parameters that do not exist for WOM codes.

While there are several constructions of WOM codes, we focus here on the one called \emph{linear WOM codes}~\cite{CGM86} in which a binary matrix is used to encode messages by the syndromes of parity check matrices of error-correcting codes. The authors of~\cite{CGM86} studied this linear construction using Golay codes as well as simplex codes. In particular, the latter family of codes provided WOM codes with the parameters $[2^r-1,r,2^{r-2}+2]$. Later, this result has been improved by Godlewski~\cite{G87}, who showed the existence of $[2^r-1,r,2^{r-2}+2^{r-4}+1]$ WOM codes.

The family of parallel RIO codes is very similar to the one of functional batch codes. In fact, if parallel RIO codes are constructed using linear codes and their parity check matrices, such as in~\cite{CGM86,G87}, then these codes are in essence functional batch codes as well. This approach to construct parallel RIO codes has been initiated recently by Yamawaki, Kamabe, and Lu in~\cite{YKL17}, where they studied the parameters of parallel RIO codes using simplex codes and showed the construction of $(7,3,4)$ and $(15,4,8)$ parallel RIO codes. These codes assure also that $FB(3,4) =7$ and  $FB(4,8) =15$. We also verified that a $(31,5,16)$ parallel RIO code exists which implies that $FB(5,16) =31$, while similarly to the conjecture raised in~\cite{YKL17} we also have the following conjecture.
\begin{conjecture}
\label{conj}
The $[2^r-1,r]$ simplex code is a functional $2^{r-1}$-batch code and therefore $FB(r,2^{r-1})=2^r-1$.
\end{conjecture}

%The simplex code has been analyzed in PIR code, batch code and RIO code. It is natural to look into its behaviour as a functional batch code.

%Therefore an $[n,s]$ simplex code has length $n=2^s-1$. An $[n,s]$ simplex code can be used as a (functional) $2^{s-1}$-PIR code since each request $\mathbf{v}$ can be recovered $2^{s-1}$ times, by $2^{s-1}-1$ pairs $(\mathbf{u},\mathbf{u}+\mathbf{v})$ and by $\mathbf{v}$ itself. Wang, Kiah, Cassuto and Bruck~\cite{WKC15,WKCB17} proved that it can also be used as a $2^{s-1}$-batch code, with the additional property that each recovering set is of size at most two. A natural problem is to analyze how many requests can be satisfied by the simplex code in the functional batch setting. The ultimate objective is to show the following Conjecture~\ref{conj}, which was established in~\cite{YKL17} for $s=3,~4$ and verified for $s=5$ as well.

%\begin{conjecture} \label{conj}
%The $[n,s]$ simplex code is a functional $2^{s-1}$-batch code and therefore\linebreak $FB(s,2^{s-1})=2^s-1$.\end{conjecture}

Remember that for WOM codes the message requests are received in a sequential order and each recovery set should be determined without knowing the upcoming requests. The main idea of the construction of $[2^r-1,r,2^{r-2}+2^{r-4}+1]$ WOM codes by Godlewski~ \cite{G87} with simplex codes works as follows.
%In \cite{G87} Godlewski proved that any $2^{s-2}+2^{s-4}+1$ requests can be satisfied when considering the simplex code as a WOM code, that is,
%The general idea of Godlewski is as follows:
\begin{enumerate}
  \item The first request $\bv$ is simply satisfied by using $\bv$ itself.

  \item As long as there are at least $2^{r-1}$ nonzero available vectors, each request $\bv$ can always be satisfied by finding a pair $\{\bu,\bu+\bv\}$. This process can satisfy at least $2^{r-2}$ more requests and only stops when the number of unused vectors is less than $2^{r-1}$.

  \item The key part of Godlewski's construction is that it is still possible to find recovery sets of size four unless the number of unused vectors is less than $2^{r-2}$. Thus in this process $2^{r-4}$ additional write requests can be satisfied.
\end{enumerate}
To summarize, simplex codes can be used to satisfy roughly any $\frac{5}{16}2^r$ write requests, when considered as WOM codes. Since in the functional batch setting (or in parallel RIO codes) we know all the requests in advance, it is possible to make use of this knowledge and improve upon the $2^{r-2}+2^{r-4}+1$ result. This improvement comes either from the choice of many recovery sets of size one, or from a predetermined usage of the $2^{r-2}$ remaining vectors in Godlewski's method. Namely, we prove the following theorem.
\begin{theorem}
The $[2^r-1,r]$ simplex code can be used as a functional $(2^{r-2}+2^{r-4}+\lfloor \frac{2^{r/2}}{\sqrt{24}}\rfloor)$-batch code.
\end{theorem}

\begin{IEEEproof}
Consider $\gamma =2^{r-2}+2^{r-4}+\lfloor\frac{2^{r/2}}{\sqrt{24}}\rfloor$ requests which consist of $\Delta$ distinct vectors $\{\bv_1,\dots,\bv_{\Delta}\}$. To prove that the simplex code is a $\gamma$-functional batch code,
we distinguish between the following two cases depending on the value of $\Delta$:

\noindent
{\bf Case 1:} If $\Delta\ge \frac{2^{r/2}}{\sqrt{6}}$, we use the $\Delta$ subsets of size one of the set
$\{\bv_1,\dots,\bv_{\Delta}\}$ as recovery sets of size one. For the remaining $\gamma-\Delta$ requests, we follow Godlewski's method.
The number of unused vectors is $2^{r}-1-\Delta$. Recovery sets of size two can be found until the number
of unused vectors is less than $2^{r-1}$. Hence, the number of recovery sets of size two is
$\frac{2^{r}-1-\Delta-(2^{r-1}-1)}{2}$ (if $\Delta$ is even) or $\frac{2^{r}-1-\Delta-(2^{r-1}-2)}{2}$ (if $\Delta$ is odd),
i.e., $2^{r-2}-\lfloor \frac{\Delta}{2} \rfloor$. Similarly, recovery sets of size four can
be found until the number of unused vectors is less than $2^{r-2}$, yielding $2^{r-4}$ recovery sets.
Therefore, when $\Delta\ge \frac{2^{r/2}}{\sqrt{6}}$, the simplex code satisfies
any $2^{r-2}+2^{r-4}+\Delta-\lfloor \frac{\Delta}{2} \rfloor\ge \gamma$ requests.

\noindent
{\bf Case 2:} If $\Delta < \frac{2^{r/2}}{\sqrt{6}}$, let $\bv_1$ be the vector which
is requested the largest number of times. Clearly, $\bv_1$ is requested at least $\lceil \frac{\gamma}{\Delta}\rceil$
times and the number of requests other than $\bv_1$ is at most $\gamma-\lceil \frac{\gamma}{\Delta} \rceil$ times.

Partition all the $2^r$ vectors (including the zero vector) into $2^{r-1}$ pairs of the form $\{\bu,\bu+\bv_1\}$.
The two vectors in the same pair are called conjugates of each other. A pair containing no requested
vectors is called a \emph{good} pair and the vectors lying in good pairs are called \emph{good} vectors.
The number of good vectors is then at least $2^{r}-2\Delta$.

For any $\bv_j\neq \bv_1$ which is requested an odd number of times, $\bv_j$ is considered
as a recovery set of size one. Hence, now each such $\bv_j$ is requested an even number of times.
For these requests we find recovery sets using only good vectors similarly to Godlewski's method.
Let $\{\bx,\by\}$ be a recovery set of size two for $\bv_j$, i.e., $\bv_j=\bx+\by$, where $\bx$ and $\by$ are good vectors.
$\bx$ and $\by$ are not conjugate since $\bv_j\neq \bv_1$). Hence, their conjugates form another recovery set for $\bv_j$,
i.e. $\bv_j=(\bx+\bv_1)+(\by+\bv_1)$. Similarly, whenever a recovery set of size four
for $\bv_j$ is found among the good vectors, then there are only two possibilities.
On one hand if we have $\bv_j=\bx+\by+\bz+\bw$ where no two of the four vectors $\{\bx,\by,\bz,\bw\}$ are
conjugate, then their conjugates form another recovery set
$\bv_j=(\bx+\bv_1)+(\by+\bv_1)+(\bz+\bv_1)+(\bw+\bv_1)$.
On the other hand if we have $\bv_j=\bx+\by+\bz+(\bz+\bv_1)$, then we construct another recovery set
$\bv_j=(\bx+\bv_1)+(\by+\bv_1)+\bw+(\bw+\bv_1)$, where the good pair $\{\bw,\bw+\bv_1\}$ is
chosen arbitrarily from the unused good pairs. After performing this strategy
for requests other than $\bv_1$ using the modified Godlewski's method, the
remaining good vectors will appear in pairs where each pair sums up
to $\bv_1$. These remaining good pairs will be used for recovering $\bv_1$.

To complete the proof we have to show that there exist enough recovery sets.
We distinguish between three subcases depending on the number of times $\lambda$ that $\bv_1$ is requested:

\noindent
{\bf Case 2.1:} If $\lambda\le 2^{r-3}$ times, then there are at
least $\frac{2^{r}-2\Delta-(2^{r-1}-2)}{2}=2^{r-2}-\Delta+1$ recovery sets
of size two and $2^{r-4}$ recovery sets of size four for the queries which are different from $\bv_1$.
This satisfies the requirements since the number of queries other than $\bv_1$ is upper bounded by
\begin{align*}
\gamma-\lceil \frac{\gamma}{\Delta} \rceil &\le 2^{r-2}+2^{r-4}+\lfloor \frac{2^{r/2}}{\sqrt{24}}\rfloor - \frac{2^{r-2}+2^{r-4}}{2^{r/2}/\sqrt{6}} \\
& \le 2^{r-2}+2^{r-4}+\lfloor \frac{2^{r/2}}{\sqrt{24}}\rfloor - 2^{r/2}\cdot \frac{5\sqrt{6}}{16} \\
& \le 2^{r-2}+2^{r-4}-\frac{2^{r/2}}{\sqrt{6}}\\
& \le 2^{r-2}+2^{r-4}-\Delta.
\end{align*}
Meanwhile, when this modified Godlewski's method concludes, there are still $2^{r-2}$
good vectors constituting $2^{r-3}$ pairs for recovering $\bv_1$.

\noindent
{\bf Case 2.2:} If $\lambda\ge \gamma +\Delta-2^{r-2}$, then the total number of requests different
than $\bv_1$ is $\gamma-\lambda$. Hence, the modified Godlewski's method concludes
after we choose $\gamma -\lambda$ recovery sets of size two. Initially, there are at least $2^{r-1}-\Delta$ good pairs,
among which $\gamma -\lambda$ pairs are involved in recovery sets of size two
(since in the modified Godlewski's method every two conjugate recovery sets of size two
together occupy two good pairs). Therefore, the number of remaining good pairs is
\begin{align*}
  2^{r-1}-\Delta-(\gamma-\lambda) &\ge 2^{r-1}-\frac{2^{r/2}}{\sqrt{6}}-(2^{r-2}+2^{r-4}+\lfloor \frac{2^{r/2}}{\sqrt{24}}\rfloor)+\lambda \\
   & \ge 2^{r-2}-2^{r-4}-2^{r/2}\cdot \frac{3}{2\sqrt{6}}+\lambda \ge \lambda,
\end{align*}
where the last inequality holds for $r\ge 6$. Thus, there are enough pairs to be used as recovery sets for $\bv_1$.

\noindent
{\bf Case 2.3:} If $ 2^{r-3}< \lambda < \gamma+\Delta-2^{r-2}$, then
the modified Godlewski's method concludes after we choose $2^{r-2}-\Delta$ recovery sets
of size two and $\gamma-\lambda-2^{r-2}+\Delta$ recovery sets of size four. Initially, there
are $2^{r-1}-\Delta$ good pairs, among which $2^{r-2}-\Delta$ pairs are
involved in recovery sets of size two and $2(\gamma-\lambda-2^{r-2}+\Delta)$ recovery sets are
involved in recovery sets of size four (since in the modified Godlewski's method every
two conjugate recovery sets of size two together occupy two good pairs and every
two conjugate recovery sets of size four together occupy four good pairs).
Thus, the number of remaining good pairs is
\begin{align}
  2^{r-1}-\Delta-(2^{r-2}-\Delta)-2(\gamma-\lambda-2^{r-2}+\Delta) &= \notag 2^{r-1}+2^{r-2}-2\gamma-2\Delta+2\lambda \\
  &\ge 2^{r-2}-2^{r/2}\cdot \frac{3}{\sqrt{6}}+\lambda \label{eq:sub2.3} \\
  &\ge \notag \lambda,
\end{align}
where (\ref{eq:sub2.3}) is derived by plugging the values of
$\gamma=2^{r-2}+2^{r-4}+\lfloor \frac{2^{r/2}}{\sqrt{24}}\rfloor$, $\Delta< \frac{2^{r/2}}{\sqrt{6}}$,
and $\lambda>2^{r-3}$. Finally, the last inequality holds for $r\ge 5$.
Therefore, there are enough pairs for recovering $\bv_1$.

Thus, the $[2^r-1,r]$ simplex code can satisfy any $2^{r-2}+2^{r-4}+\lfloor \frac{2^{r/2}}{\sqrt{24}}\rfloor$ requests.
\end{IEEEproof}

\section{Conclusions and Problems for Future Research}
\label{sec:conclude}

We have considered the shortest length of functional PIR and functional batch codes.
Several upper bounds, based on explicit constructions and random ones, are given.
Several methods which yield lower bounds are also presented.
In particular connections to WOM codes and RIO codes are derived and the
parameters of the simplex code when used as a functional batch code are discussed.

There are plenty of problems which remain for future research, some of them are
briefly outlined.
\begin{enumerate}
\item Prove or disprove that for any given PIR (batch) code,
there exists a systematic PIR (batch) code with the same parameters.

\item We would like to see an upper bound on the length of functional batch codes, which
is derived from an explicit construction.

\item We would like to see more tight bounds, general, asymptotic, and for specific parameters.

\item We would like to see a proof (or a counter-example) for Conjecture~\ref{conj}, i.e.,
the $[2^r-1,r]$ simplex code is a functional $2^{r-1}$-batch code and therefore $FB(r,2^{r-1})=2^r-1$.
\end{enumerate}

\end{document}